\documentclass[final,1p,authoryear]{elsarticle}
\usepackage{amssymb}
\usepackage{amsmath}
\usepackage{multirow}
\usepackage{multicol}   
\usepackage{pdflscape}
\usepackage{makecell}  
\usepackage{rotating}   
\usepackage{subcaption} 
\usepackage{graphicx}
\usepackage{booktabs,array}
\usepackage{subcaption} 
\usepackage[colorlinks=true, urlcolor=blue, linkcolor=blue, citecolor=blue]{hyperref}
\usepackage[section]{placeins} 
\usepackage{color}
\newcommand{\aee}[1]{\textcolor{magenta}{#1}}
\renewcommand{\aee}[1]{#1}

\journal{Spatial Statistics}

\begin{document}
\begin{frontmatter}

\title{A Bayesian INLA-SPDE Approach to Spatio-Temporal Point-Grid Fusion with Change-of-Support and Misaligned Covariates} 

\author{Weiyue Zheng, Andrew Elliott, Claire Miller, Marian Scott}

\affiliation{organization={School of Mathematics and Statistics, University of Glasgow},
addressline={University Avenue}, 
city={Glasgow},
postcode={G128QQ}, 
state={Scotland},
country={United Kingdom}}

\begin{abstract}

We propose a spatio-temporal data-fusion framework for point data and gridded data with variables observed on different spatial supports. A latent Gaussian field with a Matérn-SPDE prior provides a continuous space representation, while source-specific observation operators map observations to both point measurements and gridded averages, addressing change-of-support and covariate misalignment. Additionally incorporating temporal dependence enables prediction at unknown locations and time points. Inference and prediction are performed using the Integrated Nested Laplace Approximation and the Stochastic Partial Differential Equations approach, which delivers fast computation with uncertainty quantification. Our contributions are: a hierarchical model that jointly fuses multiple data sources of the same variable under different spatial and temporal resolutions and measurement errors, and a practical implementation that incorporates misaligned covariates via the same data fusion framework allowing differing covariate supports.
We demonstrate the utility of this framework via simulations calibrated to realistic sensor densities and spatial coverage. Using the simulation framework, we explore the stability and performance of the approach with respect to the number of time points and data/covariate availability, demonstrating gains over single-source models through point and gridded data fusion. We apply our framework to soil moisture mapping in the Elliot Water catchment (Angus, Scotland). We fuse in-situ sensor data with aligned and misaligned covariates, satellite data and elevation data to produce daily high resolution maps with uncertainty. 

\end{abstract}


\end{frontmatter}

\section{Introduction}
\label{sec1}
Spatial data fusion is a critical challenge in environmental and ecological statistics, where data are often from different sources \citep{berrocal2012space}, such as point measurements (e.g., sensors) and areal or gridded data (e.g., satellite pixels, administrative units). This paper develops a spatio-temporal Integrated Nested Laplace Approximation and the Stochastic Partial Differential Equations approach (INLA-SPDE) data fusion framework that combines point data and gridded data, enabling fine resolution predictions. Methodologically, this is related to the change-of-support problem (COSP), which merges datasets observed at different supports, resolutions, and locations, as formalised by \citet{gotway2002combining}. We also address spatial misalignment, where the response and/or covariates are measured at different locations \citep{gryparis2009measurement}, which is a practically challenging issue \citep{scott2023framing}.

Fundamental work on spatial data fusion and the change-of-support problem is often referred to as Bayesian melding. Bayesian melding merges point and gridded data through a latent spatial field and introduces aggregation methods to map the latent process to the specific spatial support and accommodate support-specific measurement errors. For example, \cite{gelfand2001change} introduce a unified Bayesian approach for prediction across multiple combinations of point and block supports, applying fully Bayesian kriging and explicitly considering support aggregation and disaggregation in the likelihood. \cite{fuentes2005model} develop Bayesian melding for fusing point observations and numerical model outputs, assuming both data types arise from a shared latent Gaussian process with explicit bias correction and uncertainty propagation. \cite{wikle2005combining} propose hierarchical conditioning to combine information across spatial scales, and \cite{gotway2007geostatistical} present a geostatistical framework compatible with GIS implementations that incorporates data supports, handles covariate misalignment, and enables prediction with uncertainty. Spatio-temporal extensions include \cite{sahu2010fusing}. For scalability, \cite{nguyen2012spatial} introduce SSDF using Fixed Rank Kriging (FRK) within an SRE model to provide low-rank covariance representations. Related developments consider mismatched spatial and temporal scales and flexible latent-field constructions \citep{wilkie2019nonparametric, wang2019combining}.

Building upon this foundation, many recent studies obtain advanced computational efficiency and flexibility by using the INLA-SPDE approach that represents the latent field via Gaussian Markov random fields on triangulated meshes \citep{lindgren2011explicit}. \cite{moraga2017geostatistical} introduce a geostatistical model for point-level and area-level spatial data within the INLA-SPDE framework, computing basis-function integrals over grid supports so both point and gridded data inform the latent field. \cite{he2024spatio} introduce the time dimension to the data fusion model based upon the work of \cite{moraga2017geostatistical} that combines in-situ sensor data and satellite data. However, they model the covariate as fixed effects, which doesn't allow for covariate misalignment. \cite{wilson2020pointless} use SPDE-based continuous spatial models to introduce area-level random effects while incorporating point data. Related work includes methodological improvements and applications in different fields \citep{suen2025cohering,zhong2023bayesian,cameletti2019bayesian,forlani2020joint,villejo2025data}.

Covariate misalignment, where the response and covariates are observed at different spatial locations (and possibly on different supports), is common in the real world. Spatial regression models \citep{cressie2015statistics} often assume the response and covariates are observed at the same locations, which is usually not true in the real world. Many approaches cannot accommodate misaligned covariates and instead require both covariates and response to be measured at the same locations and times, or they treat all data types as point-level, ignoring areal characteristics. Nearest-neighbour alignment underestimates parameter variability. The krige-and-regress (KNR) method \citep{szpiro2011efficient} aligns covariates to the response via kriging and uses Monte Carlo or bootstrap corrections for coefficient variance \citep{madsen2008regression,szpiro2011efficient,pouliot2023spatial}, but typically considers a single misaligned covariate and assumes linear relationships.

Spatio-temporal latent field models further motivate fusion under misalignment. \cite{cameletti2013spatio} develop a hierarchical spatio-temporal model with a continuously indexed Gaussian field (Matérn covariance) and autoregressive (AR) temporal dynamics, implemented efficiently for prediction and parameter estimation. Earlier spatio-temporal downscaling linked point and gridded processes without misaligned covariates \citep{mcmillan2010combining}. \cite{krainski2018advanced} show a joint model that allows spatial misalignment between response and covariate, but only in space and not in time.

To the best of our knowledge, many data fusion methods exist, but a key gap remains: misaligned covariates are typically treated as fixed effects or aligned to the response, with no original support for modelling or uncertainty propagation in space and time. We fill this gap with a Bayesian hierarchical framework with an INLA-SPDE approach that (i) models covariates as latent spatio-temporal fields on their own supports, (ii) maps both responses and covariates to observation supports (point or gridded data) via explicit aggregation operators in the likelihood, and (iii) sets different measurement errors for point sensors and gridded satellite data which reflect the fact that the two data sources have different noise characteristics. 

Our motivating application is soil moisture mapping in the Elliot Water catchment (Angus, Scotland). We fuse in-situ sensor measurements with gridded satellite products while accounting for covariate misalignment and change of support. The in-situ monitoring network includes 10 rainfall gauges and 22 soil sensors (soil temperature and soil moisture), and the satellite product provides gridded soil moisture estimates. This setting directly inspires our simulation design. The paper is structured as follows. Section \ref{sec:methodology} describes the spatio-temporal data fusion model developed to address the misaligned covariates and the change-of-support problem, as well as the INLA-SPDE approach used for fitting and inferring the proposed model. Section \ref{sec:sim_study} presents the simulation study used to validate the data fusion model under different scenarios motivated by the real dataset. Section \ref{sec:real_data_application} presents the real-world application using soil moisture data from the Elliot Water catchment in Scotland, UK. Section \ref{sec:conclusion} concludes the paper with the key findings of this study, along with the reflections and future work.

\section{Methodology}
\label{sec:methodology}

\subsection{The framework of Integrated Nested Laplace Approximation (INLA)}

INLA focuses on models that can be expressed as latent Gaussian Markov random fields (GMRF). The INLA framework can be described as follows:
$\mathbf{y}=\left(y_1, \ldots, y_n\right)$ is a vector of observed variables whose distribution is in the exponential family, and the mean $\mu_i$ (for observation $y_i$) is linked to the linear predictor $\eta_i$ using an appropriate link function. The linear predictor can include fixed effects and different random effects. $\mathbf{X}$ denotes the matrix of all latent effects which include the linear predictor, coefficients, and the distribution of the vector of latent effects is assumed to be Gaussian Markov random field (GMRF) with a zero mean and precision matrix $\mathbf{Q}\left(\boldsymbol{\theta}_2\right)$, with $\boldsymbol{\theta}_2$ a vector of hyperparameters. The distribution of $\mathbf{y}$ will depend on some vector of hyperparameters $\boldsymbol{\theta}_1$. The vectors of all hyperparameters in the model will be denoted by $\boldsymbol{\theta}=\left(\boldsymbol{\theta}_1, \boldsymbol{\theta}_2\right)$.

Observations are assumed to be independent, given the latent effects and the hyperparameters, which means the likelihood can be written as 

\begin{equation*}
    \pi(\mathbf{y} \mid \mathbf{x}, \boldsymbol{\theta})=\prod_{i \in \mathcal{I}} \pi\left(y_i \mid \eta_i, \boldsymbol{\theta}\right),
\end{equation*}

\noindent where set $\mathcal{I}$ contains indices for all observed values of $\mathbf{y}$. 

The joint posterior distribution of the effects and hyperparameters can be expressed as:

\begin{equation*}
\begin{aligned}
\pi(\mathbf{x}, \boldsymbol{\theta} \mid \mathbf{y}) &\propto \pi(\boldsymbol{\theta}) \pi(\mathbf{x} \mid \boldsymbol{\theta}) \prod_{i \in \mathcal{I}} \pi\left(y_i \mid x_i, \boldsymbol{\theta}\right) \\
&\propto \pi(\boldsymbol{\theta})|\mathbf{Q}(\boldsymbol{\theta})|^{1 / 2} \exp \left\{-\frac{1}{2} \mathbf{x}^{\top} \mathbf{Q}(\boldsymbol{\theta}) \mathbf{x}+\sum_{i \in \mathcal{I}} \log \left(\pi\left(y_i \mid x_i, \boldsymbol{\theta}\right)\right)\right\},
\end{aligned}
\end{equation*}

\noindent where $\mathbf{Q}(\boldsymbol{\theta})$ to represent the precision matrix of the latent effects. 

The marginal distributions for the latent effects and hyperparameters can be calculated from:

$$
\boldsymbol{\pi}\left(x_i \mid \mathbf{y}\right)=\int \boldsymbol{\pi}\left(x_i \mid \boldsymbol{\theta}, \mathbf{y}\right) \pi(\boldsymbol{\theta} \mid \mathbf{y}) d \theta,
$$
and
$$
\boldsymbol{\pi}\left(\theta_j \mid \mathbf{y}\right)=\int \boldsymbol{\pi}(\boldsymbol{\theta} \mid \mathbf{y}) d \theta_{-j}.
$$

Since both of the marginal distributions include integration over the space of the hyperparameters, and the dimension of $\boldsymbol{\theta}$ depends on the number of observations, which means that numerical integration is difficult for high dimensional data, a good approximation of the joint posterior distribution of the hyperparameters is required. \cite{rue2009approximate} approximate $\boldsymbol{\pi}(\boldsymbol{\theta} \mid \mathbf{y})$, denoted by $\boldsymbol{\tilde{\pi}}(\boldsymbol{\theta} \mid \mathbf{y})$, which is achieved by using the computational properties of GMRF and the Laplace approximation for multidimensional integration, and use this to approximate the posterior marginal of the latent parameter $x_i$ as:
$$
\boldsymbol{\tilde{\pi}}\left(x_i \mid \mathbf{y}\right)=\sum_k \boldsymbol{\tilde{\pi}}\left(x_i \mid \theta_k, \mathbf{y}\right) \times \boldsymbol{\tilde{\pi}}\left(\theta_k \mid \mathbf{y}\right) \times \Delta_k,
$$
\noindent where $\Delta_k$ are the weights associated with a vector of values $\theta_k$ of the hyperparameters in a grid. 

\subsection{SPDE approach} \label{sec:SPDE}
\cite{lindgren2011explicit} consider a stochastic partial differential equation (SPDE) whose solution is a Gaussian field (GF) with Matérn correlation and proposes a new approach to represent a GF with Matérn covariance, as a GMRF, by representing a solution of SPDE using the finite element method. The benefit is that the GMRF representation of the GF, which can be computed explicitly, provides a sparse representation of the spatial effect through a sparse precision matrix, which enables the nice computational properties of the GMRF, which can then be implemented in the INLA approach. To be specific, GMRF is a discrete approximation of a Gaussian field. It is obtained by discretising the continuous domain into a grid or lattice of points. In a GMRF, the values at each grid point are assumed to be conditionally independent of all other points, given their neighbouring points. This conditional independence property is often represented using a sparse precision matrix (also known as an inverse covariance matrix), where nonzero entries indicate dependencies between neighbouring points. GMRFs provide a computationally efficient way to model and analyse large spatial datasets.

The linear fractional SPDE can be defined as:

\begin{equation}
\begin{aligned}
        \left(\kappa^2-\Delta\right)^{\alpha / 2}u(\mathbf{x})=\mathbf{W}(\mathbf{x}), \quad \mathbf{x} \in \mathbb{R}^d, \quad \alpha=\nu+d / 2, \quad \kappa>0, \quad \nu>0,
        \label{eql: SPDE} 
\end{aligned}
\end{equation}

\noindent where $d$ denotes the dimension of the spatial domain, $u(\mathbf{x})$ denotes the latent field, $\Delta$ is the Laplacian operator: $\Delta = \sum_{i=1}^{d} \frac{\partial ^2}{\partial {x_i}^2}$ and $\mathbf{W}(\mathbf{x})$ denotes a spatial white noise Gaussian stochastic process with unit variance. 

Given $n$ observations $y_i, i=1, \ldots, n$, at locations $\mathbf{x}_i$, the following model can be defined:

\begin{align}
\mathbf{y} \mid \beta_0, \mathbf{u}, \sigma_e^2 & \sim N\left(\beta_0+\mathbf{A} \mathbf{u}, \sigma_e^2\right), \\
\mathbf{u} & \sim GF(0, \Sigma),
\end{align}

\noindent where $\beta_0$ is the intercept, $\mathbf{A}$ is the projection matrix and $\mathbf{u}$ is a spatial Gaussian random field. Note that the projection matrix $\mathbf{A}$ links the spatial Gaussian random field (defined using the mesh nodes, which are similar to the integration points on a numeric integration algorithm) to the locations of observed data. 

\subsection{Gaussian random field process}

A Gaussian field (GF) process can be denoted by $u(\mathbf{x})$, where $\mathbf{x}$ is any location in a study area $\mathbf{D}$. $u(\mathbf{x})$ is a stochastic process, with $\mathbf{x} \in \mathbf{D}$, where $\mathbf{D} \subset \mathbb{R}^d$. For example, $\mathbf{D}$ is a domain and data have been collected at geographical locations, over $d=2$ dimensions within this domain. The continuously indexed GF is assumed to be continuous over space and implies that it is possible to collect data at any finite set of locations within the domain. To complete the specification of the distribution of $u(\mathbf{x})$, it is necessary to define its mean and covariance. A very simple way to define a correlation function is based only on the Euclidean distance between locations, which assumes that when two pairs of points are equally distant from each other, they will exhibit an equivalent level of correlation. Matérn covariance is another widely used way to define the correlation function, with further details in the next section. 

In many scenarios, it is commonly assumed that there exists an underlying GF that cannot be directly observed. Instead, observations are data with a measurement error $e_i$,
$$
y\left(\mathbf{x}_i\right)=u\left(\mathbf{x}_i\right)+e_i,
$$
 \noindent where $e_i$ is independent of $e_j$ for all $i \neq j$ and $e_i$ follows a Gaussian distribution with zero mean and variance $\sigma_e^2$. The covariance of the marginal distribution of $y(\mathbf{x})$ at a finite number of locations is $\Sigma_y=\Sigma+\sigma_e^2 \mathbf{I}$.
 
\subsection{The Matérn covariance}

The Matérn covariance is widely used in various scientific fields to define the covariance function $\Sigma$, and the reason it is used here is that GF $u{(\mathbf{x})}$ with the Matérn covariance is a solution to the linear fractional SPDE shown in Equation \eqref{eql: SPDE}.

For two locations $\mathbf{x}_i$ and $\mathbf{x}_j$, the stationary and isotropic Matérn correlation function is defined as:
\begin{equation}
\begin{aligned}
 \operatorname{Cor}_M\left(u\left(\mathbf{x}_i\right), u\left(\mathbf{x}_j\right)\right)=\frac{1}{2^{\nu-1}\Gamma(\nu)}\left(\kappa\left\|\mathbf{x}_i-\mathbf{x}_j\right\|\right)^\nu K_\nu\left(\kappa\left\|\mathbf{x}_i-\mathbf{x}_j\right\|\right),
\end{aligned}
\label{eq:maternSpatial}
\end{equation}

\noindent where $\|$.$\|$  denotes the Euclidean distance and $K_\nu$ is the modified Bessel function of the second kind and $\nu$ is the order. To be specific, the modified Bessel function of the second kind is the function $K_{\nu}(z)$ with $z=\kappa\|x_i-x_j\|$, which is one of the solutions to the modified Bessel differential equation. $\kappa$ is a scaling parameter, which can also be interpreted as a range parameter $\rho$, representing the Euclidean distance at which $x_i$ and $x_j$ become almost independent. The empirically derived definition $\rho = \sqrt{0.8\nu}/\kappa$, corresponds to correlation near 0.1 at the distance $\rho$, for all $\nu$.

The Matérn covariance function is $\sigma_u^2 \operatorname{Cor}_M\left(u\left(\mathbf{x}_i\right), u\left(\mathbf{x}_j\right)\right)$, where $\sigma_u^2$ is the marginal variance of the process and is defined as:
$$
\sigma_u^2 = \frac{\Gamma(\nu)}{\Gamma(\nu + d/2)(4\pi)^{d/2}\kappa^{2\nu}}.
$$

\subsection{Spatio-only data fusion model framework}

\aee{Next, we will construct our fusion model. First, we will define a} 

spatio-only data fusion model built upon a geostatistical framework, 
\aee{before extending this framework to the full spatial temporal framework.}
The model assumes that there is a spatially continuous variable underlying all observations that can be modelled using a Gaussian random field process \citep{moraga2017geostatistical,ZhengIWSM2024}. Let $D$ denote the set of points with real number coordinates in a two-dimensional plane. The process is denoted by $\mu=\left\{\mu(\mathbf{s}): \mathbf{s} \in D \subset \mathbb{R}^2\right\}$, has mean function $\mathbb{E}[\mu(\mathbf{s})]=0$ and stationary covariance function $\operatorname{Cov}\left(\mu(\mathbf{s}), \mu\left(\mathbf{s}^{\prime}\right)\right)=\Sigma\left(\mathbf{s}-\mathbf{s}^{\prime}\right)$. Conditionally on $\mu(\mathbf{s})$, point data $Y_{i}$ observed at a finite set of sites, say $\mathbf{s}_{i} \in D, i= (1,\ldots,I)$, are mutually independent with

$$
Y\left(\mathbf{s}_{i}\right) \mid \mu\left(\mathbf{s}_{i}\right) \sim N\left(x\left(\mathbf{s}_{i}\right)+\mu\left(\mathbf{s}_{i}\right), \tau^2\right),
$$

\noindent where $x\left(\mathbf{s}_{i}\right)$ represents the large-scale structure of the spatial process, capturing broad variation across the study catchment. Rather than assuming a constant mean, $x\left(\mathbf{s}_{i}\right)$ allows the expected value of the response variable to vary with location $\mathbf{s}_{i}$, reflecting the influence of spatially distributed covariates such as elevation and soil type. It can be interpreted as a smoothly varying surface that describes the average behaviour of the response variable across space using a mean function, while the remaining spatial random field $\mu\left(\mathbf{s}_{i}\right)$ accounts for variation not explained by the mean function. The $\tau$ represents the standard deviation, measuring how much the observations are spread out around the mean.

The geostatistical model framework can be defined as follows for point data and grid data, respectively. Point data observations in location $\mathbf{s}_{i}, i = (1,2, \ldots, I)$. Gridded data observations are defined as block averages in blocks $\mathbf{B}_j \subset D, j= (1,2, \ldots, J)$, while a block $\mathbf{B}_j$ is a measurable subset of $D$ with $|\mathbf{B}_j| > 0$,  over which spatial processes $x(\mathbf{s})$ and $\mu(\mathbf{s})$ are averaged \citep{moraga2017geostatistical}.

\begin{equation}
\begin{aligned}
   & Y_k(\mathbf{s}_{i})=\alpha + x(\mathbf{s}_{i}) +\mu_k\left(\mathbf{s}_{i}\right) + e_k^{(p)}(\mathbf{s}_{i}), \quad i=1, \ldots, I 
\label{model: S_fusion_point_model}  
\end{aligned}
\end{equation}

\begin{equation}
\begin{aligned}
Y_k\left(\mathbf{B}_j\right)=\left|\mathbf{B}_j\right|^{-1} \int_{\mathbf{B}_j}(\alpha_k + x(\mathbf{s})+\mu_k(\mathbf{s}))d \mathbf{s} + e_k^{(g)}(\mathbf{B}_{j})  , \quad\left|\mathbf{B}_j\right|>0,
\label{model: S_fusion_grid_model}  
\end{aligned}
\end{equation}

\noindent where $k = 1,2, \ldots, K$ denote the index for $K$ different variables (such as environmental factors) and $B_j$ denotes a block in domain $D$ and $\left|\mathbf{B}_j\right|=\int_{\mathbf{B}_j} 1 d \mathbf{s}$ denotes the area of $\mathbf{B}_j$, and $e_k^{(p)}(\mathbf{s}) \sim N (0, {\tau_{k}^{(p)}}^2)$ and $e_k^{(g)}(\mathbf{B}) \sim N (0, {\tau_{k}^{(g)}}^2)$ are uncorrelated error terms defined by Gaussian white-noise processes. \aee{Note, that the variance parameters are not shared between the point and grid/areal error terms as the errors can be structural different between the two different modalities} 

The projection matrix $\mathbf{A}$ specified in the SPDE approach is designed to deal with point data, and many past studies treat gridded data as point data by using the centroid locations of the grid, thereby overlooking the inherent characteristics of gridded data. A novel way to construct the projection matrix $\mathbf{A}$ is proposed by \cite{moraga2017geostatistical}, which specifies that a particular observation in area $\mathbf{B}$ and the process $\mathbf{\mu}$ is linked through the mean value of the random field in the entire area: $\mu(\textbf{B}) = |\mathbf{B}|^{-1}\int_{B_{j}}\mu(\mathbf{s}) d \mathbf{s}$, where $\left|\mathbf{B}\right|$ denotes the area of $\mathbf{B}$. The integral defines the theoretical relationship between the latent field $\mu(\mathbf{s})$ and the gridded observation $Y_k\left(\mathbf{B}_j\right)$, and represents the true unobserved block average over the entire area $B_j$. However, computing the integral of $\mu(\mathbf{s})$ is challenging because it is a continuous process, but the mesh vertices used here only have discrete points. In the projection matrix, each row of $\mathbf{A}$ corresponds to a particular observation in block $\mathbf{B}_j$. The elements in each row are weights assigned to mesh vertices inside $\mathbf{B}_j$, which are usually 1/H, where H is the number of vertices of the mesh in $\mathbf{B}_j$. So this approximates the integral as $
\mu(B_j) \approx \frac{1}{H} \sum_{h=1}^{H} \mu(s_h),$ where $s_h$ are vertices in $\mathbf{B}_j$. Thus, $\mathbf{A}$ acts as a numerical integration operator, converting the continuous integral into a tractable discrete average. The integral of the process in each grid is approximated by taking the average of the vertex weights in the corresponding cell.

\subsection{Spatio-temporal data fusion model framework}
Building from the spatial formulation, this section introduces a spatio-temporal data fusion framework that combines point and gridded data under change of support, explicitly handles misaligned covariates and responses, and allows different measurement errors. \cite{he2024spatio} extend a spatial-only data fusion model to a spatio-temporal data fusion model, building upon the work from \cite{moraga2017geostatistical}. However, their model framework treats covariates only as fixed effects, so misaligned covariates cannot be accommodated in the framework. We address this issue with a flexible framework which leverages multiple covariates with possibly independent spatial support by modelling selected covariates as individual latent spatio-temporal fields and mapping both responses and covariates to observation supports via aggregation operators. This explicitly handles misaligned covariates and responses, and also allows different measurement errors for point sensors and gridded data. In our model, the response is driven by its own latent field together with the latent fields of the covariates. Our spatio-temporal data fusion model framework is  built based Equation \eqref{model: S_fusion_point_model} and \eqref{model: S_fusion_grid_model} and defined as follows:

Let $D \subset \mathbb{R}^2$ denote the spatial dimension and $T \subset \mathbb{R}$ the temporal dimension. The point observations are denoted as $Y(s,t)$. The $\mu(s)$ denotes the spatio-only process, which is a Gaussian distribution with a Matérn covariance structure that is independent of time. The latent spatio-temporal process is defined as $\eta = \left\{\eta(\boldsymbol{s}, t): {s} \in D, t \in T \right\}$, with mean function $\mathbb{E}[\eta(s,t)] = 0$. We model temporal dependence in the latent field using an autoregressive process 
which for this paper we assume is
of order 1 (AR(1)) based on the characteristics of the data \citep{cressie2015statistics}.

\begin{equation}
\eta(\boldsymbol{s}, t) = a*\eta(\boldsymbol{s}, t-1) + \sqrt{1 - a^2}*  \mu(\boldsymbol{s}, t), \quad t = 2, \ldots, T,
\label{equ:st_temporal_pattern}
\end{equation}

\noindent where $a$ denotes the temporal autoregressive coefficient satisfying $|a| < 1$ to ensure stationarity, $\mu(\boldsymbol{s}, t) \sim \text{GP}(0, \Sigma(\boldsymbol{s}, \boldsymbol{s}^{\prime}))$ is a spatially correlated innovation term 
which has a Gaussian distribution with a Matérn covariance structure that is independent of time, and initial stage when $t=0$: $\eta(\boldsymbol{s}, t) \sim N\left(0, \frac{\tau^2}{1 - a^2}\right)$.
To be specific, the covariance function with a Matérn covariance structure is as follows:

\begin{equation}
\begin{aligned}
\operatorname{Cov}_M\left(\mu(\mathbf{s},t^{\prime}),\mu(\mathbf{s}^{\prime},t^{\prime})\right)=\frac{1}{2^{\nu-1}\Gamma(\nu)}\left(\kappa\left
\|\mathbf{s}-\mathbf{s}^{\prime}\right\|\right)^\nu K_\nu\left(\kappa\left\|\mathbf{s}-\mathbf{s}^{\prime}\right\|\right),
 \label{equ:st_Matérn correlation function}
\end{aligned}
\end{equation}

\noindent where $\|$.$\|$ denotes the Euclidean distance and $K_\nu$ is the modified Bessel function of the second kind and $\nu$ is the order, with the parameter and function interpretations the same/similar as in Equation \eqref{eq:maternSpatial}.

The Matérn covariance function $\operatorname{Cov}_M\left(\mu(\mathbf{s},t^{\prime}\right), \left(\mu(\mathbf{s}^{\prime},t^{\prime})\right)$ defines the dependency structure between two location values measured at the same time point. It is noted that the dependency structure does not account for temporal dependence. Specifically, the covariance between two observations at the same time point is given by

\[
\text{Cov}_M(\mu(s,t^{\prime}), \mu(s^{\prime},t^{\prime})) = \Sigma(s - s^{\prime}),
\]
while the covariance between observations at different time points is assumed to be zero:
\[
\text{Cov}_M(\mu(s,t), \mu(s^{\prime},t^{\prime})) = 0 \quad \text{for } t \ne t^{\prime}.
\]
This implies that the spatio-only latent field $\mu(s,t)$ is correlated only in the spatial dimension and independent over time. The $\text{Cov}_M$ denotes the spatial Matérn covariance, but it is noted that we model a day-specific spatial field that is independent across days, while the temporal dependence is handled by a separate term AR(1). Each variable can be assigned its own independent latent process $\eta(\boldsymbol{s}, t)$ following this structure.

\subsection{Full spatio-temporal data fusion model}

The motivation of this study is to make predictions of a variable of interest in both space and time with a set of covariates that are observed both at point locations and as areal observations, which are potentially spatially misaligned from the variable of interest.

To formalise this, suppose we have a collection of $K$ different variables (i.e. $y_{k}$) which can be observed at either point locations or at an areal (grid) level. For the $k$th variable, define the point locations where it is measured as $\mathbf{s}^{(k)}$ and the areal regions where it is measured as $\mathbf{B}^{(k)}$.

Further, suppose there is an additional variable (namely the variable of interest) which we wish to predict at unknown locations which for notational convenience we will refer to as $y_{K+1}$ with corresponding point observation points as $\mathbf{s}^{(K+1)}$ and the areal observation regions as $\mathbf{B}^{(K+1)}$. Note that due to sensor placement, it is highly likely that for some $i$ and $j$ with $i \neq j$ that $\mathbf{s}^{(i)} \cap \mathbf{s}^{(j)} \neq \emptyset$. The reverse is also true, with it being very likely (as we will see in the real data application) that there exist pairs of $i$ and $j$ such that $\mathbf{s}^{(i)} \cap \mathbf{s}^{(j)}=\emptyset$. Moreover, and crucially for the task at hand, it is possible that $\mathbf{s}^{(i)} \cap \mathbf{s}^{(K+1)}=\emptyset$, i.e. that there are covariates that are misaligned from the quantity of interest, which motivates this study.

Our framework is as follows where with a slight abuse of notation we represent $y_{k}\left(\mathbf{s}_{j}^{(k)}, t\right)$ as the observation of the $k$-th variable at it's $j$-th point location at time $t$, $y_{k}\left(\mathbf{B}_{\mathbf{j}}^{(\mathbf{k})}, t\right)$ as the areal observation of the variable $k$ at the $j$-th areal observation point at time $t$. 
Thus, for $k \in\{1, \ldots, K\}$, $j_1 \in \{1,\ldots,|\mathbf{s}^{(k)}|\}$ and $j_2 \in \{1,\ldots,|\mathbf{B}^{(k)}|\}$:

\begin{equation}
\begin{aligned}
y_{k}\left(\mathbf{s}_{j_1}^{(k)}, t\right) & =\alpha_{k}
+\eta_{k}\left(\mathbf{s}_{j_1}^{(k)}, t\right)+e_{k}^{(p)}\left(\mathbf{s}_{j_1}^{(k)}, t\right) \\
y_{k}\left(\mathbf{B}_{\mathbf{j_2}}^{(\mathbf{k})}, t\right) & =\frac{1}{\left|\mathbf{B}_{\mathbf{j_2}}^{(\mathbf{k})}\right|} \int_{\mathbf{B}^{(\mathbf{k})}_{j_2}}\left[\alpha_{k}
+\eta_{k}(\mathbf{s}, t)\right] d \mathbf{s}+e_{3}^{(g)}\left(\mathbf{B}_{\mathbf{j_2}}^{(\mathbf{k})}, t_{m}\right).
\end{aligned}
\end{equation}
\aee{where $\eta_k(s,t)$ is defined as in Equation~\ref{equ:st_temporal_pattern} and thus has an additional AR(1) parameter for each $k$ which we denote as $a_k$.} Thus, allowing us to have a fully spatial-temporal prediction of the 
$K$ covariates in question, regardless of their misalignment. Additionally we wish to consider the set of variables which we can measure at all locations, e.g. in an environmental setting elevation. Let us assume that there are $l$ such variables and that $x(s,t) \in \mathbb{R}^{l}$ is a 
vector of such variables
at location $s$ and time $t$. This then allows the construction of the most generic functional form for the prediction of our variable of interest:

\begin{equation}
\begin{aligned}
y_{K+1}\!\left(\mathbf{s}_{j}^{(K+1)}, t\right)
  &= \alpha_{K+1}
   + f\!\Bigl(x\!\left(\mathbf{s}_{j}^{(K+1)}, t\right),\eta_{1}\!\left(\mathbf{s}_{j}^{(K+1)}, t\right),\ldots,
               \eta_{K}\!\left(\mathbf{s}_{j}^{(K+1)}, t\right)\Bigr) \\
  &\quad + \eta_{K+1}\!\left(\mathbf{s}_{j}^{(K+1)}, t\right)
   + e_{K+1}^{(p)}\!\left(\mathbf{s}_{j}^{(K+1)}, t\right),
\end{aligned}
\end{equation}

\begin{equation}
\begin{aligned}
y_{K+1}\!\left(\mathbf{B}^{(K+1)}_{j}, t\right)
  &= \frac{1}{\bigl|\mathbf{B}^{(K+1)}_{j}\bigr|}
     {\int_{\mathbf{B}^{(K+1)}_{j}}}
     \Bigl[\alpha_{K+1} 
     + f\!\bigl(x(s,t), \eta_{1}(s,t),\ldots,\eta_{K}(s,t)\bigr) \\
  &\qquad\qquad + \eta_{K+1}(s,t)\Bigr]\,\mathrm{d}\mathbf{s}
     + e_{K+1}^{(g)}\!\left(\mathbf{B}_{j}^{(K+1)}, t_{m}\right),
\end{aligned}
\end{equation}

\noindent
where 
$f$ is a generic function representing the impact of each of the covariates on the quantity of interest. In practice, we assume that $f$ is linear, giving the following form that can be estimated using the INLA-SPDE framework:

\begin{equation}
\begin{aligned}
y_{K+1}\!\left(\mathbf{s}_{j}^{(K+1)}, t\right)
  &= \alpha_{K+1}
   + \theta^{T}x\!\left(\mathbf{s}_{j}^{(K+1)}, t\right)
   + \sum_{k=1}^{K} \beta_k\,\eta_k\!\left(\mathbf{s}_{j}^{(K+1)}, t\right) \\
  &\quad + \eta_{K+1}\!\left(\mathbf{s}_{j}^{(K+1)}, t\right)
   + e_{K+1}^{(p)}\!\left(\mathbf{s}_{j}^{(K+1)}, t\right).
\label{eq:st_fusion_point_model}
\end{aligned}
\end{equation}

\begin{equation}
\begin{aligned}
y_{K+1}\!\left(\mathbf{B}^{(K+1)}_{j}, t\right)
  &= \frac{1}{\bigl|\mathbf{B}^{(K+1)}_{j}\bigr|}
     \int_{\mathbf{B}^{(K+1)}_{j}}
     \bigl[\alpha_{K+1} + \theta^{T}x(s,t)
     + \sum_{k=1}^{K}\beta_k\,\eta_k(s,t) \\
  &\qquad + \eta_{K+1}(s,t)\bigr]\,\mathrm{d}\mathbf{s}
     + e_{K+1}^{(g)}\!\left(\mathbf{B}^{(K+1)}_{j}, t_{m}\right).
     \label{eq:st_fusion_grid_model}
\end{aligned}
\end{equation}
where $\theta^T \in \mathbb{R}^{l}$ is the coefficients for the linear terms corresponding to the variables observed at all locations, $\beta_1,\dots \beta_K$ are the terms for the variables with a mixture of point and areal observations, $\alpha_{K+1}$ is the constant term for the variable of interest, finally, 
$e_{K+1}^{(p)}\!\left(\mathbf{s}_{j}^{(K+1)}, t\right)$
and
$e_{K+1}^{(g)}\!\left(\mathbf{B}^{(K+1)}_{j}, t_{m}\right)$
are the error terms for the point and areal measures respectively.

In practice, the INLA-SPDE fitting step combines both point and gridded data to produce posterior means of each latent field at the mesh nodes, which are used to compute predictions at unknown locations \citep{moraga2017geostatistical}. The predicted value $\hat{y}_{K+1}(\boldsymbol{s}, t)$ in any unknown locations $\boldsymbol{s}$ at time $t$ within domain $D$ is given by,

\begin{equation}
\hat{y}_{K+1}(\boldsymbol{s}, t)
= \hat{\alpha}_{K+1}
+ %
\hat{\theta}^{T}x(\boldsymbol{s}, t)
+ \sum_{k=1}^{K}\hat{\beta}_k\, \hat{\eta}_k(\boldsymbol{s}, t)
+ \hat{\eta}_{K+1}(\boldsymbol{s}, t).
\label{eq:stdata_fusion_prediction}
\end{equation}

Equation \eqref{eq:stdata_fusion_prediction} uses parameters estimated in the INLA-SPDE approach to construct predictions for the response variable $Y_{K+1}$. Each element of the prediction model is estimated as follows. The intercept term $\hat{\alpha}_{K+1}$ is obtained as the posterior mean of the corresponding fixed effect in the INLA model. The regression coefficients $\hat{\theta}$ of the covariate $x(\mathbf{s}, t)$ is also estimated as a fixed effect using posterior marginals from the INLA-SPDE approach. The scaling parameters $\hat{\beta}_k$, which are the weights of the contributions of the latent spatial fields $\hat{\eta_k}(\mathbf{s}, t)$, are similarly estimated as fixed effects within the hierarchical model.

The latent fields $\hat{\eta_k}(\mathbf{s},t) $, 
are modelled using the INLA-SPDE approach, which approximates the continuous spatial fields with Gaussian Markov random fields (GMRFs) defined using triangulated meshes. Their posterior means at the mesh nodes are computed during the model fitting step and then projected to the prediction locations through the basis functions of the SPDE mesh \citep{lindgren2015bayesian}. The spatial smoothness, range, and marginal variance of each latent field are treated as hyperparameters, inferred from their posterior distributions, and summarised by posterior means.

The predicted response $\hat{Y}_{K+1}(\mathbf{s},t)$ is constructed by combining the estimated intercept, the scaled covariate effect, the weighted contributions from the two latent fields, and the direct latent field representing $Y_{K+1}$. Although predictions can be generated for all variables, the focus here is on evaluating predictive performance specifically for the response variable $Y_{K+1}$. In generating predictions, the outputs from the fine-resolution maps are treated as point data.

Note that this functional form can be seen as an extension to that in \cite{he2024spatio}. In their model, covariates enter only as fixed effects, whereas our general framework allows covariates to be included either as fixed effects (with appropriate coverage) or as latent spatio-temporal fields with areal and/or point observations.

\section{Simulation study}\label{sec:sim_study}

This simulation study evaluates the predictive performance of a spatio-temporal data-fusion model that integrates sparse point measurements with gridded data, examining how the number of time points ($t$) and covariate missingness affect prediction and parameter estimation. We deal with the change-of-support problem for point and gridded data following \cite{moraga2017geostatistical}, add misaligned covariates, and extend the model to the spatio-temporal setting to borrow information across space and time. We also identify the minimum number of time points needed for reliable predictions (as measured by RMSE), providing practical guidance for deployments with sparse sensor networks. To assess the predictive performance of the spatio-temporal data fusion model, we conduct three simulation studies: (i) a spatial-prediction study that predicts values at unobserved locations on the final day of training; (ii) a spatio-temporal prediction study that predicts at unobserved locations one day ahead (outside the training period), testing temporal interpolation and spatial generalisation; and (iii) a robustness study that evaluates performance under realistic patterns of missingness in the gridded covariates. Together, these analyses provide a comprehensive assessment of the model's ability to fuse sparse point sensor data with satellite gridded data while maintaining accuracy across different settings.

In the simulation design, we consider three variables: $y_1$, $y_2$, and $y_3$. The variable $y_3$ represents the response, while $y_1$ and $y_2$ represent covariates \aee{(and thus $K=2$)}. Among the covariates, $y_1$ is spatially misaligned with the response, observed at $10$ independent point locations  (i.e. $\mathbf{s}^{(1)} \cap \mathbf{s}^{(3)} = \emptyset$), whereas $y_2$ is aligned and observed at the same $22$ point locations  
as $y_3$ (i.e. $\mathbf{s}^{(2)} = \mathbf{s}^{(3)} $). To mimic real data settings, each variable can be generated and observed as either point data or gridded data. The generation of spatio-temporal data in the simulation study follows the methodology in Section~\ref{sec:methodology}.
We generate latent spatial fields as Matérn GRFs using 100 independent seeds. For each seed, we build datasets with $t\in\{3,7,10\}$ time points using the same underlying realisations, so differences come from the number of time points $t$ rather than spatial randomness. We then introduce temporal correlation by specifying an separate AR(1) process for each variable. We generate point and gridded observations by adding independent Gaussian measurement noise to the latent fields, with different variances to reflect their measurement errors. For gridded data, we average $10{,}000$ field estimations within each cell to approximate gridded means. We then include a large-scale trend covariate $x(s,t)=0.2\,\text{Easting}+0.3\,\text{Northing}$ with $\theta_1=-0.2$. For each simulation, we uniformly sample 22 locations without replacement and set up two test scenarios for $y_3$: 20 held-out locations on the final training day and the same 20 locations one day ahead. The true parameter values $(\alpha,\beta,\theta,\rho,\sigma^2,a)$ are given in Table~\ref{tab:st_simulation_parameter} and are chosen based on both previous studies and real data characteristics to make sure that they are both theoretically reliable and practically feasible. 
See \ref{app1} for full implementation details including 
Figure~\ref {fig:st_flow_process} which illustrates the workflow. 

Overall, the simulation study gives a substantial workload and a comprehensive assessment of fusion performance for point and gridded data under different spatial and temporal conditions.

\newlength{\paramwidth}
\newlength{\valuewidth}
\newlength{\cellwidth}
\setlength{\paramwidth}{1.2cm}
\setlength{\valuewidth}{1cm}
\setlength{\cellwidth}{0.5cm}

\begin{table}[htbp]

\caption{Posterior summaries of intercept, scaling and variance parameters 
for the spatio-temporal fusion model (Equation \eqref{eq:st_fusion_point_model} and \eqref{eq:st_fusion_grid_model})
in the simulation study when varying both the number of available time points ($t \in \{3,7,10\}$) and dataset availability (only point (sensor) dataset/only the grid (areal) data/joint (both) datasets). 
Posteriors are summarised using the average:  posterior mean, 2.5\% posterior quantile  ($Q_{2.5\%}$) 97.5\% posterior quantile, ($Q_{97.5\%}$), and posterior RMSE over the 100 replicates in the %
study. 
}
\label{tab:parameter_estimation_comparison_1}

\begin{subtable}{\textwidth}
\centering
\small
\caption{Intercept parameters ($\alpha$)}
\begin{tabular}{ c|@{\hspace{1mm}}c@{\hspace{1mm}}|@{\hspace{1mm}}c@{\hspace{1mm}}c@{\hspace{1mm}}c@{\hspace{1mm}}c@{\hspace{1mm}}c@{\hspace{1mm}}c@{\hspace{1mm}}c@{\hspace{1mm}}c@{\hspace{1mm}}c@{\hspace{1mm}}c@{\hspace{1mm}}c@{\hspace{1mm}}c@{\hspace{1mm}}}

\toprule
\multirow{2}{*}{Param}
  & \multirow{2}{*}{True}
  & \multicolumn{4}{c|}{$t=3$}
  & \multicolumn{4}{c|}{$t=7$}
  & \multicolumn{4}{c}{$t=10$}
\\
\cmidrule(lr){3-6} \cmidrule(lr){7-10} \cmidrule(lr){11-14}
  & Value
  & Mean & $Q_{2.5\%}$ & $Q_{97.5\%}$ & RMSE
  & Mean & $Q_{2.5\%}$ & $Q_{97.5\%}$ & RMSE
  & Mean & $Q_{2.5\%}$ & $Q_{97.5\%}$ & RMSE
\\
\midrule
$\alpha_1$ point
  & 0.50
  & 0.48 & $-$0.40 & 1.34 & 0.42
  & 0.45 & $-$0.14 & 0.99 & 0.29
  & 0.47 & 0.00 & 0.94 & 0.24
\\
$\alpha_1$ grid
  & 0.50
  & 0.48 & $-$0.28 & 1.19 & 0.31
  & 0.47 & $-$0.07 & 0.94 & 0.21
  & 0.49 & 0.04 & 0.88 & 0.18
\\
$\alpha_1$ joint
  & 0.50
  & 0.48 & $-$0.28 & 1.19 & 0.31
  & 0.47 & $-$0.10 & 0.97 & 0.21
  & 0.49 & 0.01 & 0.90 & 0.19
\\
\midrule
$\alpha_2$ point
  & 0.80
  & 0.84 & 0.49 & 1.10 & 0.18
  & 0.79 & 0.57 & 1.01 & 0.10
  & 0.79 & 0.61 & 1.00 & 0.08
\\
$\alpha_2$ grid
  & 0.80
  & 0.84 & 0.48 & 1.05 & 0.15
  & 0.78 & 0.56 & 0.99 & 0.10
  & 0.80 & 0.60 & 0.98 & 0.08
\\
$\alpha_2$ joint
  & 0.80
  & 0.84 & 0.52 & 1.02 & 0.15
  & 0.79 & 0.55 & 0.98 & 0.09
  & 0.80 & 0.57 & 1.02 & 0.07
\\
\midrule
$\alpha_3$ point
  & 1.00
  & 1.34 & 0.89 & 2.06 & 0.42
  & 1.45 & 1.06 & 1.96 & 0.49
  & 1.44 & 1.17 & 1.90 & 0.47
\\
$\alpha_3$ grid
  & 1.00
  & 1.36 & 0.47 & 1.52 & 0.40
  & 1.43 & 0.66 & 1.53 & 0.46
  & 1.43 & 0.79 & 1.52 & 0.45
\\
$\alpha_3$ joint
  & 1.00
  & 1.38 & 0.69 & 1.62 & 0.43
  & 1.45 & 0.91 & 1.68 & 0.47
  & 1.44 & 1.00 & 1.65 & 0.46
\\
\bottomrule
\end{tabular}
\end{subtable}

\vspace{1em}

\begin{subtable}{\textwidth}
\centering
\small
\caption{Scaling parameters ($\beta$ \& $\theta$)}
\begin{tabular}{ c|@{\hspace{1mm}}c@{\hspace{1mm}}|@{\hspace{1mm}}c@{\hspace{1mm}}c@{\hspace{1mm}}c@{\hspace{1mm}}c@{\hspace{1mm}}c@{\hspace{1mm}}c@{\hspace{1mm}}c@{\hspace{1mm}}c@{\hspace{1mm}}c@{\hspace{1mm}}c@{\hspace{1mm}}c@{\hspace{1mm}}c@{\hspace{1mm}}}

\toprule
\multirow{2}{*}{Param}
  & \multirow{2}{*}{True}
  & \multicolumn{4}{c|}{$t=3$}
  & \multicolumn{4}{c|}{$t=7$}
  & \multicolumn{4}{c}{$t=10$}
\\
\cmidrule(lr){3-6} \cmidrule(lr){7-10} \cmidrule(lr){11-14}
  & Value
  & Mean & $Q_{2.5\%}$ & $Q_{97.5\%}$ & RMSE
  & Mean & $Q_{2.5\%}$ & $Q_{97.5\%}$ & RMSE
  & Mean & $Q_{2.5\%}$ & $Q_{97.5\%}$ & RMSE
\\
\midrule
$\beta_1$ point
  & $-0.30$
  & $-$0.21 & $-$0.47 & $-$0.08 & 0.14
  & $-$0.21 & $-$0.39 & $-$0.17 & 0.04
  & $-$0.18 & $-$0.37 & $-$0.16 & 0.04
\\
$\beta_1$ grid
  & $-0.30$
  & $-$0.28 & $-$0.38 & $-$0.12 & 0.08
  & $-$0.18 & $-$0.37 & $-$0.21 & 0.03
  & $-$0.15 & $-$0.37 & $-$0.23 & 0.02
\\
$\beta_1$ joint
  & $-0.30$
  & $-$0.28 & $-$0.34 & $-$0.12 & 0.07
  & $-$0.16 & $-$0.34 & $-$0.11 & 0.03
  & $-$0.13 & $-$0.33 & $-$0.10 & 0.02
\\
\midrule
$\beta_2$ point
  & $-0.40$
  & $-$0.31 & $-$0.54 & 0.01 & 0.15
  & $-$0.34 & $-$0.50 & $-$0.16 & 0.08
  & $-$0.27 & $-$0.46 & $-$0.18 & 0.06
\\
$\beta_2$ grid
  & $-0.40$
  & $-$0.33 & $-$0.55 & $-$0.01 & 0.14
  & $-$0.40 & $-$0.56 & $-$0.18 & 0.10
  & $-$0.33 & $-$0.56 & $-$0.24 & 0.08
\\
$\beta_2$ joint
  & $-0.40$
  & $-$0.36 & $-$0.73 & $-$0.06 & 0.11
  & $-$0.42 & $-$0.70 & $-$0.29 & 0.10
  & $-$0.33 & $-$0.66 & $-$0.24 & 0.10
\\
\midrule
$\theta_1$ point
  & $-0.20$
  & $-$0.19 & $-$0.57 & 0.02 & 0.13
  & $-$0.41 & $-$0.49 & $-$0.03 & 0.12
  & $-$0.35 & $-$0.45 & $-$0.08 & 0.07
\\
$\theta_1$ grid
  & $-0.20$
  & $-$0.19 & $-$0.25 & 0.26 & 0.09
  & $-$0.39 & $-$0.20 & 0.22 & 0.08
  & $-$0.33 & $-$0.18 & 0.18 & 0.08
\\
$\theta_1$ joint
  & $-0.20$
  & $-$0.18 & $-$0.25 & 0.19 & 0.09
  & $-$0.36 & $-$0.24 & 0.13 & 0.08
  & $-$0.29 & $-$0.23 & 0.10 & 0.06
\\
\bottomrule
\end{tabular}
\end{subtable}

\vspace{1em}

\begin{subtable}{\textwidth}
\centering
\small
\caption{Variance parameters ($\sigma^2$)}
\begin{tabular}{ c|@{\hspace{1mm}}c@{\hspace{1mm}}|@{\hspace{1mm}}c@{\hspace{1mm}}c@{\hspace{1mm}}c@{\hspace{1mm}}c@{\hspace{1mm}}c@{\hspace{1mm}}c@{\hspace{1mm}}c@{\hspace{1mm}}c@{\hspace{1mm}}c@{\hspace{1mm}}c@{\hspace{1mm}}c@{\hspace{1mm}}c@{\hspace{1mm}}}
\toprule
\multirow{2}{*}{Param}
  & \multirow{2}{*}{True}
  & \multicolumn{4}{c|}{$t=3$}
  & \multicolumn{4}{c|}{$t=7$}
  & \multicolumn{4}{c}{$t=10$}
\\
\cmidrule(lr){3-6} \cmidrule(lr){7-10} \cmidrule(lr){11-14}
  & Value
  & Mean & $Q_{2.5\%}$ & $Q_{97.5\%}$ & RMSE
  & Mean & $Q_{2.5\%}$ & $Q_{97.5\%}$ & RMSE
  & Mean & $Q_{2.5\%}$ & $Q_{97.5\%}$ & RMSE
\\
\midrule
$\sigma^2_1$ point
  & 1.00
  & 1.00 & 0.72 & 1.56 & 0.16
  & 1.04 & 0.81 & 1.32 & 0.10
  & 1.06 & 0.86 & 1.29 & 0.10
\\
$\sigma^2_1$ grid
  & 1.00
  & 0.96 & 0.72 & 1.25 & 0.15
  & 0.93 & 0.78 & 1.11 & 0.13
  & 0.91 & 0.78 & 1.05 & 0.14
\\
$\sigma^2_1$ joint
  & 1.00
  & 0.92 & 0.69 & 1.21 & 0.17
  & 0.90 & 0.74 & 1.09 & 0.16
  & 0.87 & 0.74 & 1.03 & 0.16
\\
\midrule
$\sigma^2_2$ point
  & 0.50
  & 0.46 & 0.33 & 0.64 & 0.10
  & 0.48 & 0.39 & 0.58 & 0.06
  & 0.49 & 0.41 & 0.58 & 0.05
\\
$\sigma^2_2$ grid
  & 0.50
  & 0.45 & 0.32 & 0.61 & 0.07
  & 0.45 & 0.36 & 0.55 & 0.07
  & 0.44 & 0.37 & 0.53 & 0.06
\\
$\sigma^2_2$ joint
  & 0.50
  & 0.35 & 0.24 & 0.51 & 0.19
  & 0.39 & 0.31 & 0.49 & 0.15
  & 0.41 & 0.33 & 0.51 & 0.15
\\
\midrule
$\sigma^2_3$ point
  & 0.30
  & 0.32 & 0.16 & 0.61 & 0.07
  & 0.32 & 0.21 & 0.46 & 0.07
  & 0.30 & 0.22 & 0.41 & 0.04
\\
$\sigma^2_3$ grid
  & 0.30
  & 0.33 & 0.21 & 0.50 & 0.07
  & 0.30 & 0.21 & 0.55 & 0.03
  & 0.29 & 0.22 & 0.38 & 0.04
\\
$\sigma^2_3$ joint
  & 0.30
  & 0.25 & 0.14 & 0.42 & 0.09
  & 0.29 & 0.20 & 0.41 & 0.12
  & 0.28 & 0.20 & 0.39 & 0.10
\\
\bottomrule
\end{tabular}
\end{subtable}
\end{table}

\begin{table}[htbp]

\caption{Posterior summaries of range parameters and temporal coefficients 
for the spatio-temporal fusion model (Equation \eqref{eq:st_fusion_point_model} and \eqref{eq:st_fusion_grid_model})
in the simulation study when varying both the number of available time points ($t \in \{3,7,10\}$) and dataset availability (only point (sensor) dataset/only the grid (areal) data/joint (both) datasets). Posteriors are summarised using the average: posterior mean, 2.5\% posterior quantile  ($Q_{2.5\%}$) 97.5\% posterior quantile, ($Q_{97.5\%}$), and posterior RMSE over the 100 replicates in the study.}
\label{tab:parameter_estimation_comparison_2}

\ContinuedFloat
\centering

\begin{subtable}{\textwidth}
\centering
\small
\caption{Range parameters ($\rho$)}
\begin{tabular}{ c|@{\hspace{1mm}}c@{\hspace{1mm}}|@{\hspace{1mm}}c@{\hspace{1mm}}c@{\hspace{1mm}}c@{\hspace{1mm}}c@{\hspace{1mm}}c@{\hspace{1mm}}c@{\hspace{1mm}}c@{\hspace{1mm}}c@{\hspace{1mm}}c@{\hspace{1mm}}c@{\hspace{1mm}}c@{\hspace{1mm}}c@{\hspace{1mm}}}

\toprule
\multirow{2}{*}{Param}
  & \multirow{2}{*}{True}
  & \multicolumn{4}{c|}{$t=3$}
  & \multicolumn{4}{c|}{$t=7$}
  & \multicolumn{4}{c}{$t=10$}
\\
\cmidrule(lr){3-6} \cmidrule(lr){7-10} \cmidrule(lr){11-14}
  & Value
  & Mean & $Q_{2.5\%}$ & $Q_{97.5\%}$ & RMSE
  & Mean & $Q_{2.5\%}$ & $Q_{97.5\%}$ & RMSE
  & Mean & $Q_{2.5\%}$ & $Q_{97.5\%}$ & RMSE
\\
\midrule
$\rho_1$ point
  & 4.00
  & 4.91 & 2.25 & 9.59 & 1.74
  & 4.31 & 2.76 & 6.43 & 1.14
  & 4.06 & 2.82 & 5.65 & 0.81
\\
$\rho_1$ grid
  & 4.00
  & 4.87 & 3.04 & 7.45 & 1.10
  & 4.78 & 3.54 & 6.34 & 0.94
  & 4.78 & 3.70 & 6.09 & 0.90
\\
$\rho_1$ joint
  & 4.00
  & 5.03 & 3.14 & 7.73 & 1.26
  & 5.11 & 3.76 & 6.84 & 1.22
  & 5.08 & 3.92 & 6.52 & 1.17
\\
\midrule
$\rho_2$ point
  & 3.00
  & 3.84 & 1.75 & 7.71 & 1.74
  & 3.12 & 2.05 & 4.55 & 0.39
  & 3.06 & 2.14 & 4.22 & 0.30
\\
$\rho_2$ grid
  & 3.00
  & 3.70 & 1.91 & 6.64 & 1.24
  & 3.81 & 2.55 & 5.55 & 0.88
  & 3.82 & 2.73 & 5.23 & 0.90
\\
$\rho_2$ joint
  & 3.00
  & 4.57 & 2.14 & 8.93 & 2.52
  & 3.83 & 2.45 & 5.76 & 1.42
  & 3.89 & 2.69 & 5.47 & 1.20
\\
\midrule
$\rho_3$ point
  & 2.00
  & 3.31 & 1.14 & 17.85 & 1.73
  & 3.00 & 1.63 & 6.27 & 1.48
  & 2.60 & 1.59 & 4.56 & 0.87
\\
$\rho_3$ grid
  & 2.00
  & 3.39 & 2.16 & 11.30 & 1.65
  & 3.12 & 2.68 & 8.60 & 1.31
  & 2.95 & 2.55 & 6.93 & 1.06
\\
$\rho_3$ joint
  & 2.00
  & 2.82 & 1.27 & 18.79 & 1.05
  & 2.62 & 2.04 & 8.19 & 0.80
  & 2.58 & 1.88 & 6.18 & 0.74
\\
\bottomrule
\end{tabular}
\end{subtable}

\vspace{1em}

\begin{subtable}{\textwidth}
\centering
\small
\caption{Temporal coefficients ($a$)}
\begin{tabular}{ c|@{\hspace{1mm}}c@{\hspace{1mm}}|@{\hspace{1mm}}c@{\hspace{1mm}}c@{\hspace{1mm}}c@{\hspace{1mm}}c@{\hspace{1mm}}c@{\hspace{1mm}}c@{\hspace{1mm}}c@{\hspace{1mm}}c@{\hspace{1mm}}c@{\hspace{1mm}}c@{\hspace{1mm}}c@{\hspace{1mm}}c@{\hspace{1mm}}}

\toprule
\multirow{2}{*}{Param}
  & \multirow{2}{*}{True}
  & \multicolumn{4}{c|}{$t=3$}
  & \multicolumn{4}{c|}{$t=7$}
  & \multicolumn{4}{c}{$t=10$}
\\
\cmidrule(lr){3-6} \cmidrule(lr){7-10} \cmidrule(lr){11-14}
  & Value
  & Mean & $Q_{2.5\%}$ & $Q_{97.5\%}$ & RMSE
  & Mean & $Q_{2.5\%}$ & $Q_{97.5\%}$ & RMSE
  & Mean & $Q_{2.5\%}$ & $Q_{97.5\%}$ & RMSE
\\
\midrule
$a_1$ point
  & 0.40
  & 0.34 & $-$0.15 & 0.73 & 0.27
  & 0.33 & 0.05 & 0.58 & 0.20
  & 0.34 & 0.10 & 0.56 & 0.17
\\
$a_1$ grid
  & 0.40
  & 0.43 & 0.12 & 0.68 & 0.15
  & 0.39 & 0.21 & 0.56 & 0.06
  & 0.39 & 0.24 & 0.53 & 0.05
\\
$a_1$ joint
  & 0.40
  & 0.45 & 0.13 & 0.70 & 0.14
  & 0.44 & 0.24 & 0.61 & 0.08
  & 0.44 & 0.27 & 0.58 & 0.05
\\
\midrule
$a_2$ point
  & 0.50
  & 0.40 & $-$0.03 & 0.72 & 0.19
  & 0.45 & 0.24 & 0.62 & 0.13
  & 0.46 & 0.29 & 0.61 & 0.11
\\
$a_2$ grid
  & 0.50
  & 0.42 & 0.04 & 0.71 & 0.26
  & 0.44 & 0.20 & 0.65 & 0.11
  & 0.46 & 0.26 & 0.63 & 0.07
\\
$a_2$ joint
  & 0.50
  & 0.47 & 0.04 & 0.82 & 0.37
  & 0.60 & 0.40 & 0.77 & 0.20
  & 0.64 & 0.48 & 0.78 & 0.20
\\
\midrule
$a_3$ point
  & 0.60
  & 0.49 & $-$0.16 & 0.89 & 0.31
  & 0.59 & 0.17 & 0.81 & 0.15
  & 0.59 & 0.26 & 0.81 & 0.16
\\
$a_3$ grid
  & 0.60
  & 0.44 & 0.10 & 0.89 & 0.40
  & 0.57 & 0.53 & 0.92 & 0.15
  & 0.59 & 0.57 & 0.90 & 0.14
\\
$a_3$ joint
  & 0.60
  & 0.59 & 0.14 & 0.97 & 0.21
  & 0.61 & 0.56 & 0.96 & 0.10
  & 0.62 & 0.73 & 0.95 & 0.07
\\
\bottomrule
\end{tabular}
\end{subtable}

\end{table}

\subsection{Final-day predictions at unobserved locations on the last day of the training period with varying time points.} 

We compare the prediction performance and parameter estimation of the point-only, grid-only, and joint (fusion) models for final-day predictions at unobserved locations, varying the number of time points $t$. Table \ref{tab:parameter_estimation_comparison_1} and \ref{tab:parameter_estimation_comparison_2} presents the parameter estimation performance with the RMSE and mean of the spatio-temporal point, grid, and joint models across varying numbers of time points ($t$) on simulation datasets. Most parameter RMSE values decrease with increasing $t$, suggesting improved estimation accuracy. Notably, the joint model consistently outperforms both the point and grid models, with lower RMSE values in most scenarios. This performance shows the joint model's ability to outperform the other two models in the spatio-temporal scenario, effectively borrowing information across both dimensions. In contrast, the point and grid models exhibit less accurate estimated parameters, which suggests that fusing different data sources improves the parameter estimation.

The RMSE improvement is very obvious as $t$ increases from 3 to 7 time points, suggesting that temporal information enhances estimation by reducing uncertainty through repeated measurements on the same locations. However, beyond $t=7$,  the improvement stabilises, with little difference observed between $t=7$ and $t=10$. This plateau implies a threshold where additional temporal data adds limited information for parameter estimation, perhaps because the model has already captured much of the temporal structure. From a real data application perspective, this finding suggests that allocating resources to increase temporal resolution beyond $t=10$ may not guarantee better prediction performance. Instead, refining spatial resolution or incorporating more data sources could be more helpful for further model improvements.

The first step to validate the spatio-temporal data fusion model is to assess its predictive performance at the time boundary of the training period. In this simulation study, the whole dataset spans 100 days, with the final day used as the test set. From this day, 20 unobserved locations are randomly selected as test points. The training set consists of the last $t$ days, including the final day, and is used to train the model. Predictions are made for the test points on the last day. We predict the final-day test points. INLA-SPDE models spatial structure via an SPDE (a GMRF approximation to a Matérn field) and temporal dependence via an AR(1) process. Predictions are calculated using the posterior mean of the latent field, and performance is measured by root mean squared prediction error (RMSPE).

\begin{figure}
         \centering
         \includegraphics[trim={0cm 0cm 0cm 0cm},clip,width=\textwidth]{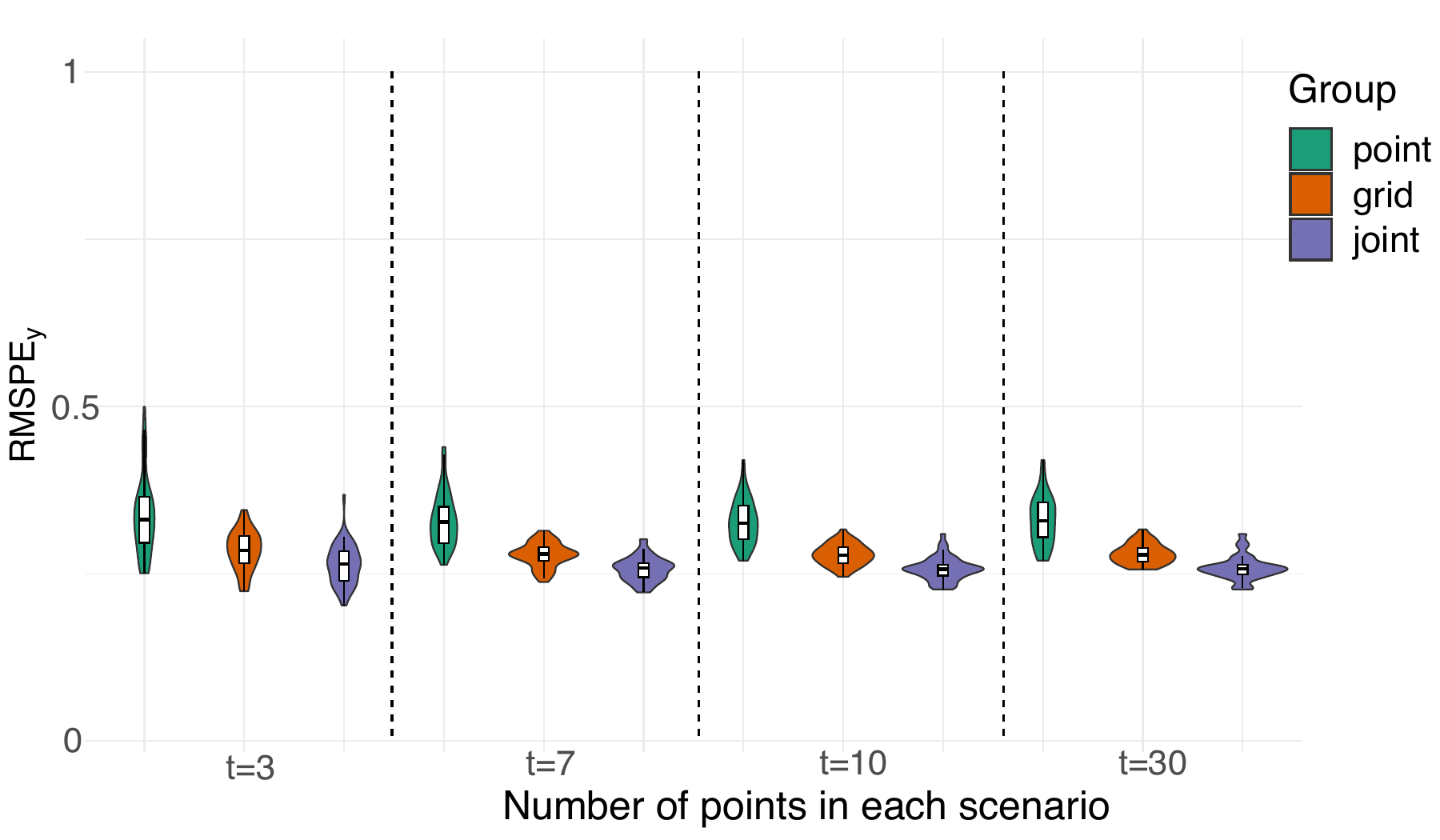}
         \caption{Comparison of prediction error ($\text{RMSPE}_y$) at unobserved locations of point, grid, and joint models with varying numbers of time points ($t \in \{ 3, 7, 10, 30\}$). Results based on 100 simulations with latent field ($\sigma_1 = 1, \sigma_2 = 0.5, \sigma_3 = 0.3$) of final day predictions of the last day of the training period.}
\label{fig:st_prediction_last_day_training_set}
\end{figure}

Figure \ref{fig:st_prediction_last_day_training_set} compares predictive accuracy for the point, grid, and joint models at unobserved locations on the final day of the training period. The violin plots show prediction errors ($\text{RMSPE}_y$) across 100 simulations, with lower values indicating better performance. As the number of time points increases, the joint model consistently gets the lowest $\text{RMSPE}_y$, while the point model shows very little improvement. The spread of the joint model's violin plot also decreases, indicating less across-replicate spread in the simulations. These all together demonstrate that the joint model is more effective at using temporal information to reduce uncertainty when compared to the point and grid models.

\subsection{One-day-ahead future predictions at unobserved locations beyond the training period with varying time points ($t = 3,7,10,30$).} 

\begin{figure}
         \centering
         \includegraphics[trim={0cm 0cm 0cm 0cm},clip,width=\textwidth]{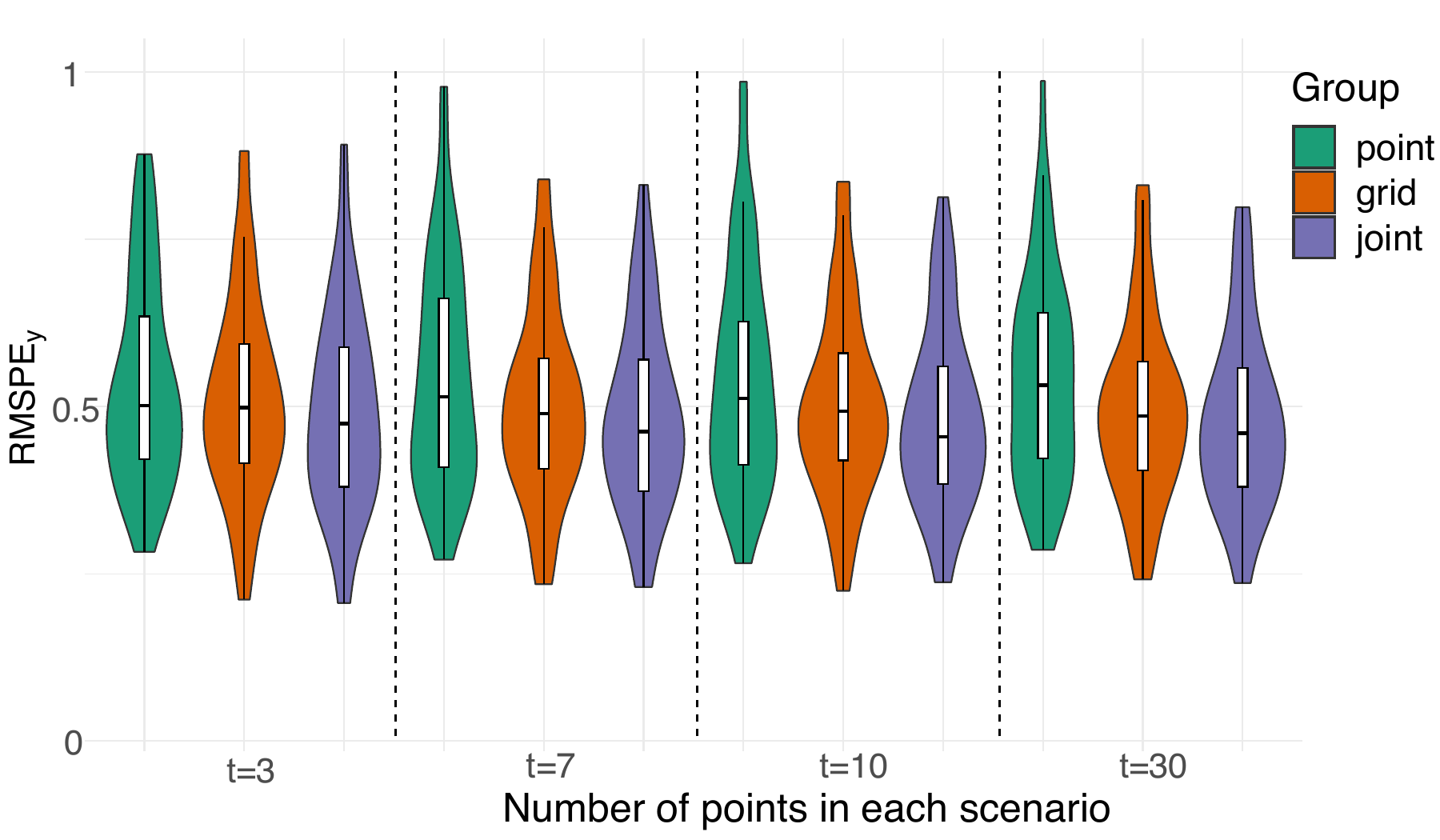}
         \caption{Comparison of prediction error ($\text{RMSPE}_y$) at unobserved locations of point, grid, and joint models with varying numbers of time points ($ t \in \{3, 7, 10, 30\}$). Results based on 100 simulations with latent field ($\sigma_1 = 1, \sigma_2 = 0.5, \sigma_3 = 0.3$) of one-day-ahead predictions of the training period.}
         \label{fig:st_prediction_one_more_day_training_set}
\end{figure}

In this simulation study, the dataset includes 100 days, with the final day used as the test set. On this day, 20 unobserved locations are randomly selected as test points. The training set consists of the t days immediately before the final day (it excludes the last day) and is used to fit the model. Predictions are then made for the test points on the final day.

Figure \ref{fig:st_prediction_one_more_day_training_set} shows the one-day-ahead prediction errors ($\text{RMSPE}_y$) at unobserved locations. The joint model still outperforms or performs equally well as the point and grid model, though with smaller margins than in Figure \ref{fig:st_prediction_last_day_training_set}. All models show higher overall error (0.4-0.6 range) for future predictions beyond the training period, with performance differences remaining relatively stable despite increasing time points. The distributions of the $\text{RMSPE}_y$ appear wider, suggesting more variability in one-day-ahead prediction accuracy. This suggests that while temporal information improves model performance, predicting future time points at unobserved locations remains more challenging.

\subsubsection{Assessing the model performance under realistic grid covariate missingness}

\begin{table}[ht]
\centering
\caption{Posterior summaries of parameters 
for the spatio-temporal fusion joint model (Equation \eqref{eq:st_fusion_point_model} and \eqref{eq:st_fusion_grid_model})
in the simulation study 
varying the availability/missingness of gridded covariates.
Experiments are conducted with $t=3$.
Posteriors are summarised using the average:  posterior mean, 2.5\% posterior quantile  ($Q_{2.5\%}$) 97.5\% posterior quantile, ($Q_{97.5\%}$), and posterior RMSE over the 100 replicates in the study. 
}
\label{fig:missing_covariates_comparison}
\begin{tabular}{l c 
                c c c c | c c c c}
\toprule
& & \multicolumn{4}{c|}{Gridded covariates missing} 
  & \multicolumn{4}{c}{Gridded covariates complete} \\
\cmidrule(lr){3-6}\cmidrule(lr){7-10}
Param
& True 
  & Mean & Q$_{2.5\%}$ & Q$_{97.5\%}$ & RMSE 
  & Mean & Q$_{2.5\%}$ & Q$_{97.5\%}$ & RMSE \\
\midrule
$\alpha_{1}$ joint  
  & 0.50 & 0.44 & $-$0.42 & 1.29 & 0.36 
         & 0.48 & $-$0.10 & 1.24 & 0.31 \\

$\alpha_{2}$ joint  
  & 0.80 & 0.79 & 0.47 & 1.10 & 0.15 
         & 0.84 & 0.61 & 1.02 & 0.15 \\

$\alpha_{3}$ joint  
  & 1.00 & 0.98 & 0.40 & 1.56 & 0.24 
         & 1.38 & 0.69 & 1.62 & 0.43 \\

\addlinespace
$\beta_{1}$ joint   
  & $-0.30$ & $-$0.16 & $-$0.34 & 0.02 & 0.17 
         & $-$0.28 & $-$0.34 & $-$0.12 & 0.07 \\

$\beta_{2}$ joint   
  & $-0.40$ & $-$0.23 & $-$0.52 & 0.06 & 0.26 
         & $-$0.36 & $-$0.73 & $-$0.06 & 0.11 \\

$\beta_{3}$ joint   
  & $-0.20$ & $-$0.02 & $-$0.29 & 0.25 & 0.21 
         & $-$0.18 & $-$0.25 & 0.19 & 0.09 \\

\addlinespace
$\sigma^2_{1}$ joint 
  & 1.00 & 0.97 & 0.62 & 1.48 & 0.19 
         & 0.92 & 0.69 & 1.21 & 0.17 \\

$\sigma^2_{2}$ joint 
  & 0.50 & 0.46 & 0.32 & 0.66 & 0.09 
         & 0.35 & 0.24 & 0.51 & 0.19 \\

$\sigma^2_{3}$ joint 
  & 0.30 & 0.37 & 0.22 & 0.58 & 0.12 
         & 0.25 & 0.14 & 0.42 & 0.09 \\

\addlinespace
$\rho_{1}$ joint 
  & 4.00 & 5.55 & 2.46 & 10.92 & 2.12 
         & 5.03 & 3.14 & 7.73 & 1.26 \\

$\rho_{2}$ joint 
  & 3.00 & 3.72 & 1.69 & 7.42 & 1.68 
         & 4.57 & 2.14 & 8.93 & 2.52 \\

$\rho_{3}$ joint 
  & 2.00 & 5.38 & 2.01 & 12.76 & 5.53 
         & 2.82 & 1.27 & 18.79 & 1.05 \\

\addlinespace
$a_{1}$ joint   
  & 0.40 & 0.33 & $-$0.24 & 0.79 & 0.26 
         & 0.45 & 0.13 & 0.70 & 0.14 \\

$a_{2}$ joint   
  & 0.50 & 0.45 & 0.02 & 0.75 & 0.24 
         & 0.47 & 0.04 & 0.82 & 0.37 \\

$a_{3}$ joint   
  & 0.60 & 0.76 & 0.35 & 0.95 & 0.20 
         & 0.58 & 0.14 & 0.97 & 0.20 \\
\bottomrule
\end{tabular}
\end{table}

In our real data application case, the only available satellite data is the response variable Soil Water Index (SWI), while all the covariates (e.g., rainfall and temperature) from satellite data are not available. To ensure the proposed model framework remains robust under this setting, a targeted simulation study was designed to mimic the real data scenario. Specifically, we compare the performance of the joint model in two situations: one \aee{where 
gridded 
data 
for 
covariates ($y_1$ and $y_2$) %
are available, and another where gridded data for these variables are completely missing.} The point data, including both responses and covariates, are kept identical across both situations.

To reflect the structure of the real application, we fix the number of time points at $t = 3$. The only difference between the two scenarios is the availability of the grid covariates. Table~\ref{fig:missing_covariates_comparison} presents the posterior summaries of the joint model parameters under two scenarios: with and without the gridded covariates. 
\aee{Each parameter estimate
is summarised using the average:  posterior mean, 2.5\% posterior quantile  ($Q_{2.5\%}$) 97.5\% posterior quantile, ($Q_{97.5\%}$), and posterior root mean squared error (RMSE) over the 100 replicates in the study. 
}

The results show that the joint model produces reasonably accurate estimates for the intercept parameters $\alpha_1$, $\alpha_2$, and $\alpha_3$ under both scenarios. However, under the gridded covariates missing condition, posterior uncertainty increases slightly, and RMSEs are higher, particularly for $\alpha_3$, which appears more sensitive to missing gridded covariates. In contrast, the scaling parameters $\beta_1$, $\beta_2$, and $\theta_1$ are more affected by missing gridded covariates. These parameters show greater posterior variability and higher RMSEs in the grid missing situation, indicating reduced identifiability when covariate information is incomplete. The spatial variance parameters $\sigma^2_1$, $\sigma^2_2$, and $\sigma^2_3$ are estimated reasonably well in both scenarios. Although credible intervals are slightly wider and RMSEs slightly higher under the grid missing situation, the estimates remain close to the true values, suggesting the model maintains robustness for these latent fields' variance terms.

However, the spatial range parameters $\rho_1$, $\rho_2$, and $\rho_3$ are poorly recovered when the gridded covariates are missing, with posterior means overestimating the true values and large RMSEs. This suggests that the range parameters are structurally difficult to identify in this setting, likely due to the fact that when gridded covariates are missing, the latent fields lose large-scale structure, which might lead to the overestimation of $\rho$.

Finally, the temporal coefficients $a_1$, $a_2$, and $a_3$ remain relatively stable between the two simulation studies. The posteriors and RMSEs change only slightly when gridded covariates are added in the joint model, suggesting inference in the joint model is dominated by the point data.

\section{Real data application}
\label{sec:real_data_application}
Soil moisture is important for hydrology, agriculture, and climate \citep{Entekhabi1996, Seneviratne2010}, and monitoring it is vital for drought and flood management, irrigation planning, and improving climate models. Yet making accurate, consistent maps over large areas is challenging. We map soil moisture in the Elliot Water catchment by combining sparse but accurate SEPA in-situ measurements with the wide coverage of Copernicus satellite data. By fusing point and gridded data, we produce fine-resolution maps with quantified uncertainty, leveraging sensor accuracy and satellite spatio-temporal coverage. We apply the models in Equation \eqref{eq:st_fusion_point_model} and \eqref{eq:st_fusion_grid_model}
to data from the Scottish Environment Protection Agency (SEPA)\footnote{\url{https://www2.sepa.org.uk/sensornet}}, Copernicus\footnote{\url{https://land.copernicus.eu/global/products/ssm}} and the Open Elevation\footnote{\url{https://open-elevation.com}}. We define three variables according to our real soil moisture data: $y_1$: rainfall (spatially misaligned covariate), $y_2$: soil temperature (spatially aligned covariate), $y_3$: volumetric water content (VWC, response variable) and $x$: elevation (the fixed effect). The in-situ monitoring network includes 10 rainfall gauges and 22 soil stations that report soil temperature and moisture. The satellite provides estimates of the soil water index (SWI), a quantity related to VWC. SWI is a derived Copernicus product \citep{copernicus_soil_moisture}, available on a daily basis at 1 km resolution, corresponding to a $5 \times 19$ grid over our study catchment. The real data application incorporates multiple days of soil moisture data, allowing for the evaluation of temporal information in prediction modelling. By including data from multiple time points, the spatio-temporal data fusion model aims to determine whether modelling temporal dependencies alongside spatial correlation leads to better predictive performance on the real datasets. Specifically, the temporal information enables the model to potentially use patterns such as soil moisture persistence, seasonal effects, or delayed responses to covariates (rainfall). This section aims to quantify the gains in prediction accuracy from the incorporation of temporal structure, thereby providing insights into the advantages of spatio-temporal modelling for soil moisture data fusion.

\begin{sidewaysfigure}
\begin{tabular}{c@{\hskip 0cm}c@{\hskip 10cm}c@{\hskip 6cm}c}
&\multicolumn{3}{c}{\includegraphics[trim={0cm 0cm 0cm 15
cm},clip,width=\textwidth]{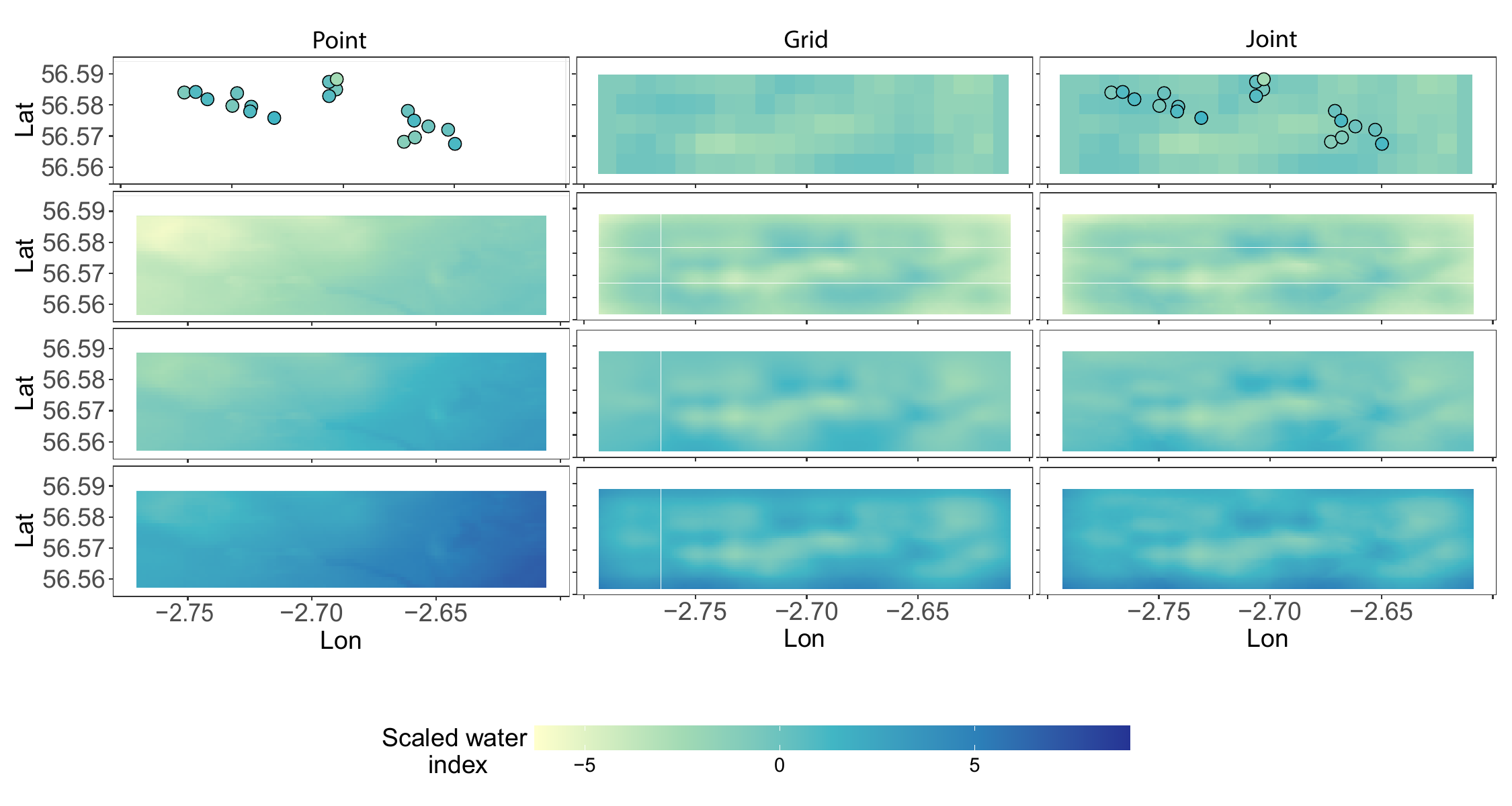}} \vspace{-7cm}\\
\begin{sideways} \hspace{0.2cm} $2.5\%$ \end{sideways} \\
\begin{sideways} \hspace{0.2cm}  Mean \end{sideways} \\
\begin{sideways} \hspace{0.2cm} $97.5\%$ \end{sideways}  
\end{tabular}
 \vspace{3cm}
\caption{Prediction maps of the standardised water index (computed from VWC and SWI) over the Elliott water catchment area on 16/05/2022 (the 10-day training set includes data from 06/05/2022 to 15/05/2022), using the same mesh, with 95\% confidence intervals. The top row displays the real data for each model, while the subsequent rows show the 95\%  prediction interval, including the 2.5\% and 97.5\% quantiles as well as the mean of the predictions.}
\label{fig:st_real_pred_map}
\end{sidewaysfigure}

Figure \ref{fig:st_real_pred_map} presents the one-day-ahead prediction map based on a 10-day training set from 06/05/2022 to 15/05/2022. We construct a standardised water index by standardising in-situ VWC measurements and satellite SWI values so that they are on a common scale. It is noted that the gridded covariates of the satellite data are missing, so the only available gridded data is the response variable. It displays the raw observations, the 95\% prediction interval, and the mean predictions for the point model, grid model and joint model. It reveals that the spatial pattern is strongly dominated by elevation in the point model, the grid model captures more spatial detail due to its broader spatial coverage, and the joint model, which integrates both point and gridded data, narrows the CIs of the prediction by using the strengths of both datasets.

\setlength{\paramwidth}{1.5cm}  
\setlength{\valuewidth}{5cm} 

\begin{table}
\caption{Parameter estimates with posterior means and 95\% CIs, obtained by fitting the point, grid and joint spatio-temporal data fusion models to the soil moisture data.}
\label{tab:st_parameter_estimates}
\begin{subtable}{\textwidth}
\centering
\caption{Intercepts}
\begin{tabular}{>{\raggedright\arraybackslash}p{\paramwidth}|>{\centering\arraybackslash}p{\valuewidth}|>{\centering\arraybackslash}p{\valuewidth}}
\toprule
Param & $t=3$ & $t=10$ \\
\midrule
$\alpha_1$ point & $-0.55\,(-0.59,\,-0.51)$ & $-0.29\,(-0.42,\,-0.16)$ \\
$\alpha_1$ joint & $-0.55\,(-0.59,\,-0.51)$ & $-0.30\,(-0.63,\,0.02)$ \\

\midrule
$\alpha_2$ point & $0.12\,(-0.34,\,0.58)$ & $0.02\,(-0.18,\,0.22)$ \\
$\alpha_2$ joint & $0.18\,(-1.07,\,1.42)$ & $0.12\,(-0.17,\,0.42)$ \\

\midrule
$\alpha_3$ point & $0.33\,(-0.92,\,1.58)$ & $0.33\,(-4.30,\,4.96)$ \\
$\alpha_3$ grid & $0.42\,(-3.90,\,4.75)$ & $0.22\,(-7.53,\,7.98)$ \\
$\alpha_3$ joint & $0.29\,(-4.18,\,4.76)$ & $0.17\,(-7.88,\,8.22)$ \\

\bottomrule
\end{tabular}
\end{subtable}

\begin{subtable}{\textwidth}
\centering
\caption{Scaling parameters}
\begin{tabular}{>{\raggedright\arraybackslash}p{\paramwidth}|>{\centering\arraybackslash}p{\valuewidth}|>{\centering\arraybackslash}p{\valuewidth}}
\toprule

$\beta_1$ point & $0.01\,(-0.55,\,0.58)$ & $0.11\,(-0.40,\,0.66)$ \\
$\beta_1$ joint & $-0.05\,(-0.50,\,0.39)$ & $2.51\,(2.18,\,2.83)$ \\
\midrule
$\beta_2$ point & $-0.07\,(-0.35,\,0.22)$ & $0.27\,(0.15,\,0.38)$ \\
$\beta_2$ joint & $-0.81\,(-1.28,\,-0.33)$ & $0.60\,(0.38,\,0.85)$ \\
\midrule
$\theta_1$ point & $-1.70\,(-2.18,\,-1.22)$ & $-1.18\,(-1.38,\,-0.98)$ \\
$\theta_1$ grid & $-0.31\,(-0.59,\,-0.04)$ & $-0.14\,(-0.52,\,0.24)$ \\
$\theta_1$ joint & $-0.85\,(-1.06,\,-0.63)$ & $-0.35\,(-0.52,\,-0.17)$ \\
\bottomrule
\end{tabular}
\end{subtable}

\begin{subtable}{\textwidth}
\centering
\caption{Spatial parameters (variance and range)}
\label{tab:spatial_params_var_range}
\begin{tabular}{>{\raggedright\arraybackslash}p{\paramwidth}|>{\centering\arraybackslash}p{\valuewidth}|>{\centering\arraybackslash}p{\valuewidth}}
\toprule

$\sigma^2_1$ point & $0.02\,(0.01,\,0.05)$ & $0.11\,(0.05,\,0.20)$ \\
$\sigma^2_1$ joint & $0.03\,(0.01,\,0.07)$ & $1.31\,(0.78,\,2.07)$ \\
\midrule

$\sigma^2_2$ point & $0.92\,(0.61,\,1.30)$ & $0.93\,(0.80,\,1.09)$ \\
$\sigma^2_2$ joint & $0.92\,(0.41,\,1.82)$ & $0.98\,(0.80,\,1.18)$ \\
\midrule

$\sigma^2_3$ point & $1.13\,(0.73,\,1.69)$ & $0.33\,(0.12,\,0.67)$ \\
$\sigma^2_3$ grid & $2.84\,(2.05,\,3.87)$ & $3.09\,(2.44,\,3.87)$ \\
$\sigma^2_3$ joint & $2.89\,(2.09,\,3.91)$ & $3.13\,(2.47,\,3.92)$ \\
\midrule
$\rho_1$ point & $3.6\times10^{4}\,\bigl(1.1\times10^{4},\,9.8\times10^{4}\bigr)$ & $4.3\times10^{4}\,\bigl(2.0\times10^{4},\,9.1\times10^{4}\bigr)$ \\
$\rho_1$ joint & $4.2\times10^{4}\,\bigl(1.3\times10^{4},\,1.2\times10^{5}\bigr)$ & $2.1\times10^{4}\,\bigl(1.5\times10^{4},\,2.8\times10^{4}\bigr)$ \\
\midrule
$\rho_2$ point & $1.8\times10^{3}\,\bigl(6.0\times10^{2},\,4.3\times10^{3}\bigr)$ & $9.2\times10^{2}\,\bigl(6.1\times10^{2},\,1.3\times10^{3}\bigr)$ \\
$\rho_2$ joint & $8.1\times10^{3}\,\bigl(2.9\times10^{3},\,1.9\times10^{4}\bigr)$ & $3.5\times10^{3}\,\bigl(2.6\times10^{3},\,4.7\times10^{3}\bigr)$ \\
\midrule
$\rho_3$ point & $7.8\times10^{3}\,\bigl(3.5\times10^{3},\,1.5\times10^{4}\bigr)$ & $8.3\times10^{4}\,\bigl(3.7\times10^{4},\,1.8\times10^{5}\bigr)$ \\
$\rho_3$ grid & $9.1\times10^{3}\,\bigl(7.5\times10^{3},\,1.1\times10^{4}\bigr)$ & $1.1\times10^{4}\,\bigl(8.7\times10^{3},\,1.3\times10^{4}\bigr)$ \\
$\rho_3$ joint & $6.2\times10^{3}\,\bigl(4.6\times10^{3},\,8.4\times10^{3}\bigr)$ & $1.1\times10^{4}\,\bigl(8.7\times10^{3},\,1.3\times10^{4}\bigr)$ \\
\bottomrule
\end{tabular}
\end{subtable}
\end{table}

\begin{table}
\ContinuedFloat
\caption{Parameter estimates with posterior means and 95\% credible intervals (2.5\% and 97.5\% quantiles), obtained by fitting the point, grid and joint spatio-temporal data-fusion models to the soil moisture data. (continued)}
\label{tab:Temporal coefficient parameters}
\begin{subtable}{\textwidth}
\centering
\caption{Temporal coefficient parameters}
\begin{tabular}{>{\raggedright\arraybackslash}p{\paramwidth}|>{\centering\arraybackslash}p{\valuewidth}|>{\centering\arraybackslash}p{\valuewidth}}
\toprule

Param & t=3 & t=10 \\
\midrule
$a_1$ point & 0.36 (-0.51, 0.92) & 0.33 (-0.20, 0.78) \\
$a_1$ joint & 0.53 (-0.46, 0.99) & 0.98 (0.94, 0.99) \\
\midrule

$a_2$ point & 0.09 (-0.33, 0.48) & 0.07 (-0.09, 0.23) \\
$a_2$ joint & 0.46 (-0.43, 0.94) & -0.22 (-0.51, 0.07) \\
\midrule

$a_3$ point & 0.05 (-0.51, 0.60) & 0.92 (0.73, 0.99) \\
$a_3$ grid & 0.98 (0.97, 0.99) & 0.98 (0.97, 0.99) \\
$a_3$ joint & 0.98 (0.97, 0.99) & 0.98 (0.97, 0.99) \\
\bottomrule
\end{tabular}
\end{subtable}
\end{table}

Table \ref{fig:missing_covariates_comparison} and \ref{tab:st_parameter_estimates} show the posterior means along with the 95\% credible intervals (2.5\% and 97.5\% quantiles) for the spatio-temporal model with different numbers of time points ($t = 3$ and $t = 10$). The number of time points corresponds to using data from $t$ days before the prediction date (16/05/2022). The parameters are grouped into intercepts, scaling parameters, spatial parameters, and temporal coefficients. Each parameter is evaluated across three models (joint, point, and grid) within the spatio-temporal model framework, with the model fitted for time points of $t=3$ and $t = 10$. The intercept estimates across the point grid and the joint model vary, and for most of the intercepts, there is less uncertainty with $t=10$ than with $ t=3$. The scaling parameters suggest that the effects are different between $t=3$ and $t=10$, with the exception that $\theta_1$ remains negative across both values of $t$, which is plausible as soil temperature or rainfall patterns likely changed between these time points. The range ($\rho$) and variance ($\sigma$) also vary a lot, and the temporal coefficients ($a_3$) indicate strong temporal autocorrelation for $y_3$.

The real data application reveals a trade-off between temporal autocorrelation and parameter uncertainty in the spatio-temporal model. While increasing the number of time points ($t=10$) improves precision for parameters with strong temporal persistence (e.g., coefficients like $a_3$), it increases uncertainty for intercepts ($\alpha_3$). The conflict reflects model structural constraints introduced by more time points (sparse daily point data, weak identifiability or stronger spatio-temporal interactions). The spatial range may vary over time, but the current model assumes it is fixed. For weakly identified parameters (e.g., $\beta_1$ and $\beta_2$), additional time points lead to unstable parameter estimation, while strongly autocorrelated temporal processes ($a_3 \approx 1$) benefited from more time points. The results highlight the importance of balancing model complexity with data adequacy: more temporal data improves signals for dominant processes but increases noise in hierarchical parameters, which needs careful prior specification or model redesign to stabilise inferences.

It is noted that increasing the time points does not guarantee decreasing the uncertainty for all parameters. Adding more time points provides more repeated measures over time, which contributes to the estimation of the temporal coefficients by capturing long-term patterns and reduces the uncertainty of the parameter estimation (e.g., narrower CIs for temporal coefficients $a_k$). In Table \ref{tab:Temporal coefficient parameters}, the $a_1$ of the joint model at $t = 10$ has a posterior mean with a tight interval compared to $t = 3$, where the CIs are wider. For example, when $t=3$, the temporal coefficient $a_1$ from the joint model is estimated at 0.53 with a 95\% CI of (-0.46, 0.99), whereas for $t = 10$, it increases to 0.98 with a narrower CI of (0.94, 0.99). This indicates that the certainty in the temporal coefficients improves as the number of time points increases in the model. In Table \ref{tab:spatial_params_var_range}, uncertainty does not always decrease with more time points. The estimation of $\rho_3$ varies and its intervals widen, whereas intervals for $\rho_1$ and $\rho_2$ tighten. This instability for $\rho_3$ likely reflects the nonstationarity for the process, so the assumption of the same daily range may be inappropriate.

\section{Discussion and conclusions}
\label{sec:conclusion}
We develop a spatio-temporal data fusion model for point and gridded data, building upon the spatial framework of \cite{moraga2017geostatistical} and its spatio-temporal extension by \cite{he2024spatio}. However, the latter's model framework treats covariates only as fixed effects, so misaligned covariates cannot be accommodated in the framework. We address this issue by modelling selected covariates as individual latent spatio-temporal fields and mapping both responses and covariates to observation supports via aggregation operators. With arbitrary point and areal supports, it explicitly handles misaligned covariates and responses, and also allows different measurement errors for point sensors and gridded data. Key structures include a latent Gaussian random field with Matérn spatial covariance extended over time via an AR(1), and a fusion strategy that integrates point sensor data and gridded satellite data under a separable spatio-temporal dependence. It makes it possible to model the environmental process from both spatial and temporal dimensions because such processes change continuously across space and through time. The spatio-temporal data fusion model is fitted using the INLA-SPDE approach for fast inference and prediction. The model assumes that the point and gridded data come from the same latent field with trend covariates, and point and gridded data are mapped using the projection matrix in the SPDE.

The simulations study the effects of $t$ on prediction and parameter estimation, robustness under missing gridded covariates, and differences between point and grid model specifications. The model is robust to gridded covariates' missingness, and performance is almost unchanged. Across point, grid, and joint models evaluated at $t \in \{3,7,10\}$, increasing $t$ generally improves the parameter estimation and prediction: RMSE decreases for $\alpha_1,\alpha_2,\beta_1,\beta_2$ (e.g., the RMSE of $\beta_1$ drops by $35\%$ from $t=3$ to $t=10$), and credible intervals narrow. The joint model consistently outperforms the point and grid model by about $15\text{-}20\%$ in RMSEs for scaling parameters and $10\text{-}15\%$ for spatial variances at $t=10$. Some biases still exist (especially in $\alpha_3$ and $\rho_3$), indicating sensitivity to modelling assumptions such as separable covariance, fixed smoothness, and the over-simplified AR(1) temporal structure. The joint model's robustness highlights the importance of spatio-temporal frameworks in complex systems with evolving spatial patterns. Overall, incorporating multiple time points improves predictive performance and enhances the accuracy and stability of parameter estimation. In particular, increasing the time points from $t=3$ to $t=10$ narrows CIs and reduces bias for fixed effects and hyperparameters because temporal replication at the same locations adds information about the latent process and measurement noise.

The spatio-temporal model improves both parameter estimation and prediction. Increasing the time points from $t=3$ to $t=10$ narrows CIs and reduces bias for fixed effects and hyperparameters, because temporal replication at the same locations adds information about the latent process and measurement noise.

In the real data application, predicted mean surfaces remain smooth far from sensors. With a sparse network, the model borrows strength from covariates. Elevation is available everywhere and explains large-scale variation, so uncertainty widens where there is little nearby sensor support. The contrast between $t=3$ and $t=10$ shows a trade-off: adding time points improves estimates and predicts even without new sensor data, but temporal replication cannot fully compensate for limited spatial coverage, and fine-scale detail still needs more locations or gridded data.

Future work can be improved in several directions. Firstly, relaxing the model structure. In particular, moving beyond separable spatio-temporal covariances (e.g., non-separable covariance or SPDEs with time varying coefficients), allows spatially varying and possibly varying ranges (treat $\rho(\mathbf{s})$ as covariate dependent or as a random field), and generalises the temporal process from AR(1) to more flexible forms such as AR($p$) or ARMA. Secondly, temporal misalignment should be incorporated in a unified way. The model could be extended to handle irregular temporal sampling and different temporal support for the point and gridded data. Thirdly, future work should systematically assess robustness to irregular sampling by evaluating performance under spatial and temporal irregularity (e.g., long gaps, informative missingness) via blocked spatio-temporal cross-validation. Finally, data quality is very important, as the current monitoring network in the real data is sparse and unevenly distributed, optimised placement of the sensors, together with additional covariates, is expected to improve the identifiability of $(\rho,\sigma^2)$ and reduce bias.

\section*{Acknowledgements}

We thank Armando Marino and Claire Neil for helpful discussions and for providing us with soil moisture data. This paper has been prepared using the European Union's Copernicus Land Monitoring Service information \url{https://land.copernicus.eu/global/products/ssm}  and the in-situ sensor data from Scottish Environment Protection Agency (SEPA) \url{https://www2.sepa.org.uk/sensornet}. W. Zheng acknowledges funding from the Chinese Scholarship Council and the fee waiver from the University of Glasgow.

\section*{Code}
The code used for the simulation study and real data application in the paper is publicly available at \href{https://github.com/weiyue-zheng/STDF-COSP}{STDF-COSP on GitHub}.

\clearpage
\newpage

\appendix

\section{Appendix 1: Additional details about the simulation study} 
\label{app1}

\begin{figure}[htbp]
         \centering
         \includegraphics[trim={0cm 0cm 4cm 0cm},clip,width=\textwidth]{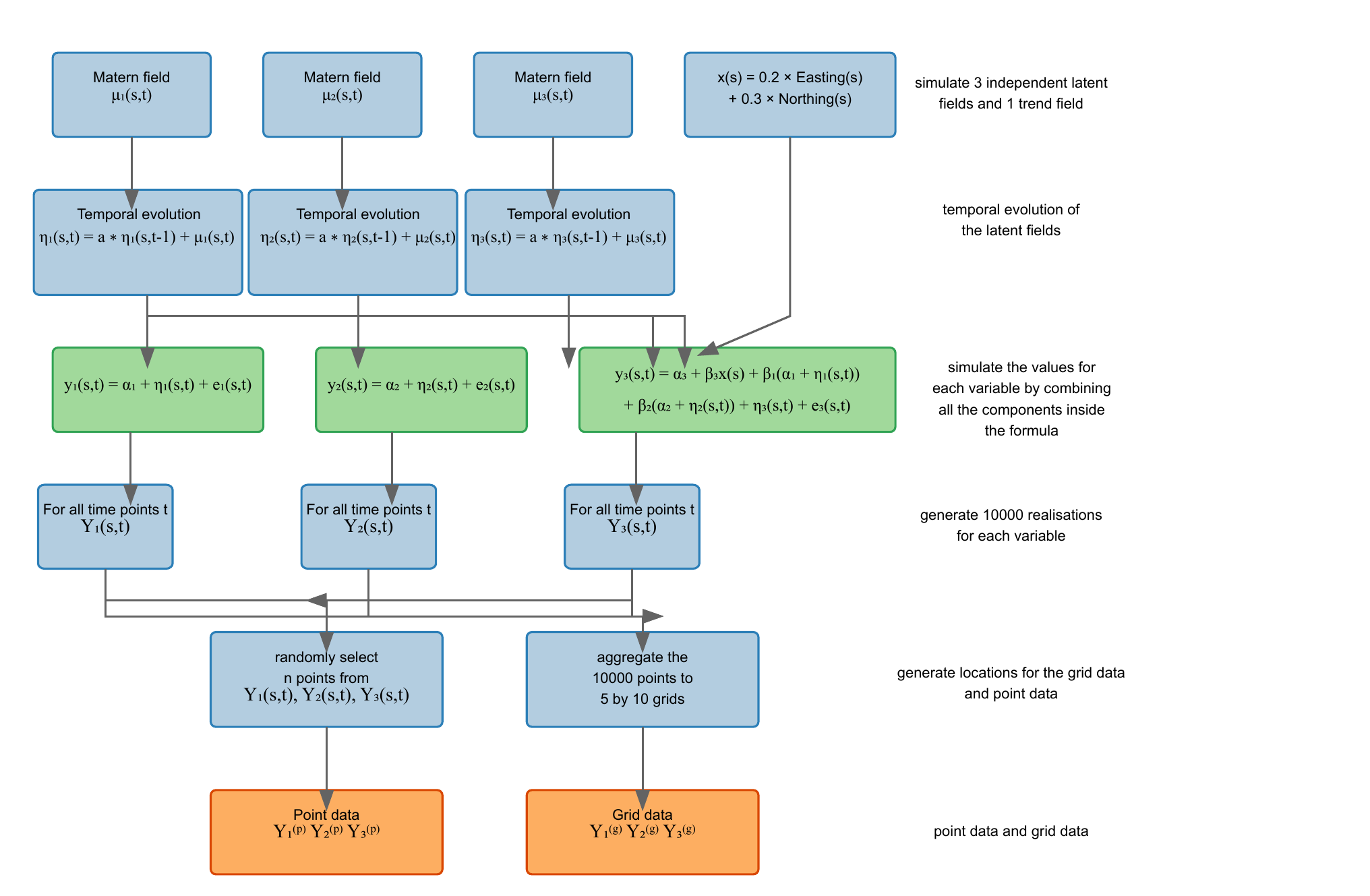}
         \caption{Flowchart illustrating the simulation process for the spatio-temporal study. The first row shows the generation of independent latent fields for each variable. The second row introduces temporal dependence. The third row combines these components into a full spatio-temporal realisation. Subsequent rows demonstrate how gridded data and point data are derived from this realisation.}
         \label{fig:st_flow_process}
\end{figure}

\begin{enumerate}
     \item Spatial processes $\mu(s)$ are simulated by generating 100 independent random field realisations from a Matérn Gaussian random field as fixed seeds, and the same 100 seeds are reused when constructing datasets of length $t$ days. Holding the spatial field fixed over time makes sure that differences across scenarios arise from $t$, not from spatial variability. The behaviour of the Matérn Gaussian field is controlled through three parameters within the Matérn covariance function: range ($\rho$), marginal variance ($\sigma$), and smoothness ($v$). 

     \item The temporal correlation is introduced by the formula as follows:
     $$\eta(s,t) = a*\eta(s,t-1) + \sqrt{1-a^2}*\mu(s,t),$$
    where the $\sqrt{1-a^2}$ term is used to make the process stationary in time. The spatio-temporal process is assumed to be a series of GRFs, and the latent spatial processes $\mu_k(s)$ generated from the first step account for the temporal dependencies through AR(1).

    \item The trend covariate $x(s,t)$, which represents the geological trend of the study catchment, is derived from a surface where values exhibit an increasing pattern from the southwest to the northeast (from 0 to 3.5) across the whole study catchment. Let the coordinates of $y(s_i,t)$ be denoted by $\text{Easting}_i$ and $\text{Northing}_i$, then the trend is defined as follows:
    
    \begin{equation}
            x(s_i,t) = 0.2*\text{Easting}_i + 0.3*\text{Northing}_i
    \end{equation}
    
    \item The uncorrected measurement error terms for point data ($e_p$) and gridded data ($e_g$) are generated from a Gaussian white-noise process: $N(0,\tau^2_{p})$ and $N(0,\tau^2_{g})$.
    \item Then the covariates and the response variables are generated by combining the previously constructed terms based on Equations \eqref{eq:st_fusion_point_model} and \eqref{eq:st_fusion_grid_model}.

Table \ref{tab:st_simulation_parameter} shows the true parameters used in the spatio-temporal simulation study. The parameters for the simulation study are chosen based on both previous studies and real data characteristics to make sure that they are both theoretically reliable and practically feasible. Some parameters, such as the intercepts $\alpha$ and precision parameters $\tau$, are borrowed from previous studies to maintain comparison with other models. Others, such as scaling parameters $\beta$, spatial parameters $\rho$ and $\sigma$, and the temporal coefficients $a$, are chosen from real data applications to reflect spatial patterns and characteristics in the real soil moisture dataset. This allows the simulation to balance theoretical evidence with real data conditions, which makes the assessment of the model's performance meaningful.

\begin{table}
    \centering
        \caption{True parameter values used in the spatio-temporal simulation data}
    \label{tab:st_simulation_parameter}
    \renewcommand{\arraystretch}{2}
    \setlength{\tabcolsep}{2pt}
    \begin{tabular}{c|ccccccccccccccccccccc}
        \hline
         &$\alpha_1$ & $\alpha_2$ & $\alpha_3$ & $\beta_1$ & $\beta_2$ & $\theta_1$ & $\rho_1$ & $\rho_2$ & $\rho_3$ & $\sigma_1$ & $\sigma_2$ & $\sigma_3$ & $\tau^2_{p1}$ & $\tau^2_{p2}$ & $\tau^2_{p3}$& $\tau^2_{g1}$ & $\tau^2_{g2}$ & $\tau^2_{g3}$ & $a_1$ & $a_2$ & $a_3$ \\
         \hline
         \multirow{1}{*}{\makecell{True\\values}}& 0.5 & 0.8 & 1 & -0.3 & -0.4 & -0.2 & 4 & 3 & 2 & 1 & 0.5 & 0.3 & 0.09 & 0.04 & 0.01 & 0.25 & 0.16 & 0.09 & 0.4 & 0.5 & 0.6 \\
        \hline
    \end{tabular}
\end{table}

    \item The gridded data is generated by first simulating independent realisations of a Matérn Gaussian random field to model the latent fields. Then, for each time point, values for gridded cells are calculated by averaging all points within that cell to ensure that each gridded cell represents the localised mean of the specific latent field. The process can be defined as: $Y^{(g)}\left(\mathbf{B}\right) = \frac{1}{n} \sum_{i=1}^{n} y_i$, where $Y^{(g)}\left(\mathbf{B}\right)$ represents the average value of the gridded cell, which indicates the mean of all values $y_i$ within the gridded cell, $n$ denotes the total number of points within the gridded cell $\mathbf{B}$, and $y_i$ represents the value of the $i$th points within the gridded cell in each day.
    
    \item To assess the model's ability to generalise unobserved data, the test set includes 20 randomly selected unobserved point locations for the response variable $y_3$ on the final day of the training period and one day ahead of the training period, with test points always at the same locations across the point model, grid model and joint mode within each simulation. Randomly selecting test locations across different simulations gives a comprehensive evaluation of the model's out-of-sample performance, reducing potential bias and assessing how well the fusion model predicts at unobserved locations.
\end{enumerate}

This section characterises the spatio-temporal simulation data used in the spatio-temporal simulation study, which provides insight into how spatial patterns evolve over time.

\begin{figure}
         \centering
         \includegraphics[trim={0cm 7cm 0cm 1cm},clip,width=\textwidth]{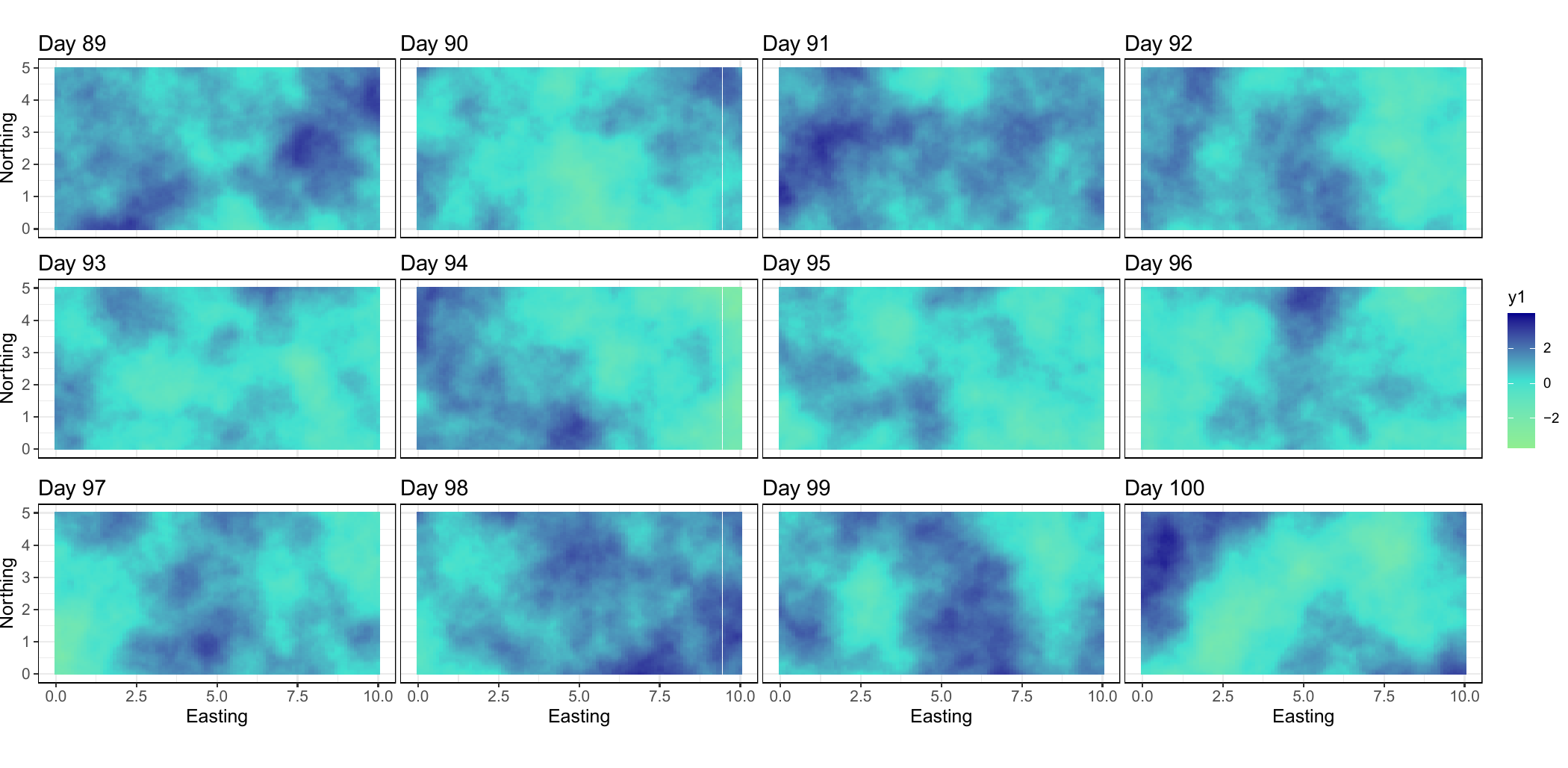}
         \caption{Final 12-day sequence of simulated latent fields for simulated rainfall $y_1$ in the spatio-temporal simulation study, using a latent field configuration ($\sigma_1 = 1$, $\sigma_2 = 0.5$, $\sigma_3 = 0.3$) and a temporal coefficient $ a_1 = 0.4$.}
         \label{fig:st_latent_fields_y_1}
\end{figure}

\begin{figure}
         \centering
         \includegraphics[trim={0cm 7cm 0cm 1cm},clip,width=\textwidth]{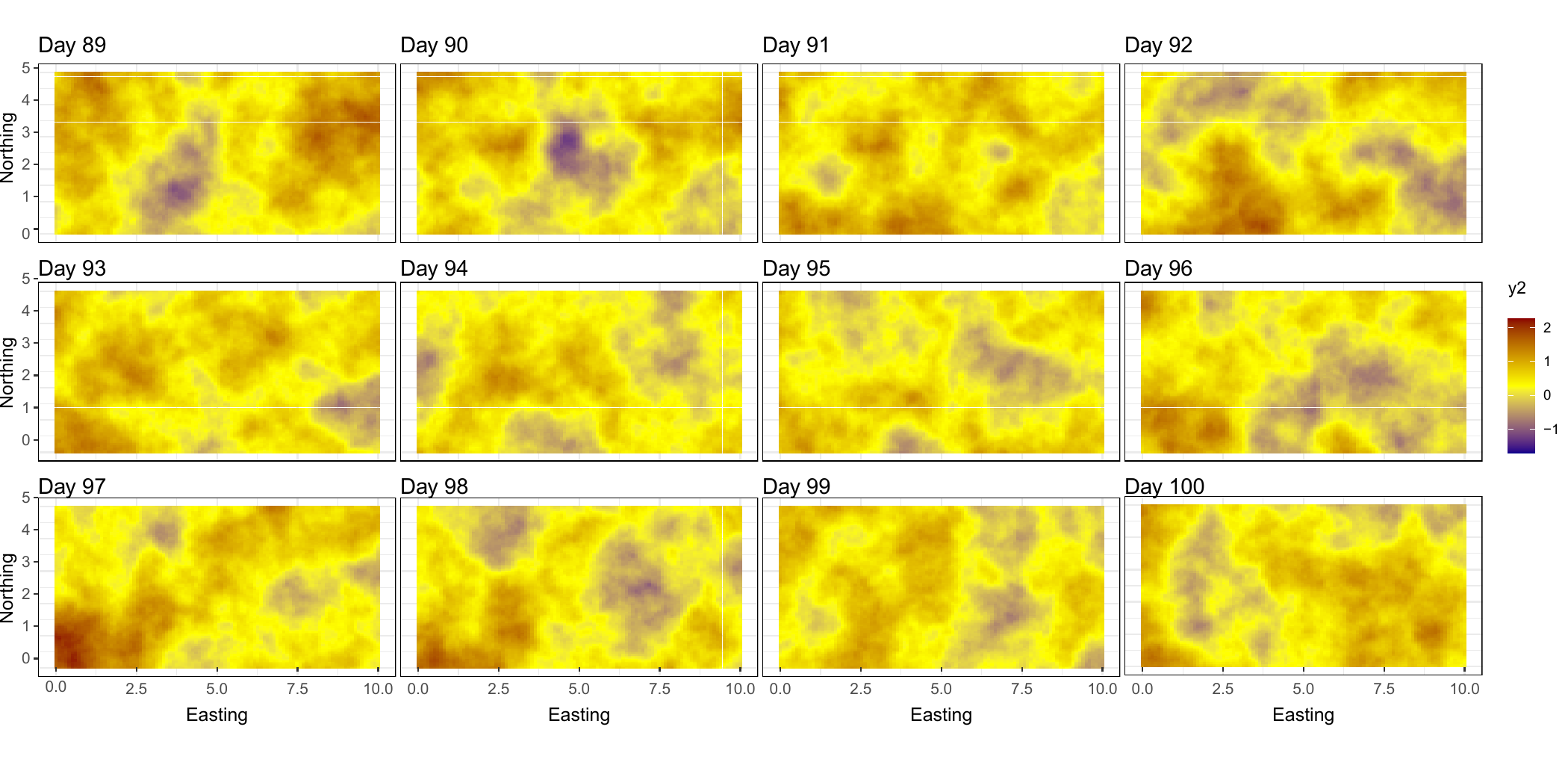}
         \caption{Final 12-day sequence of simulated latent fields for simulated soil temperature $y_2$ in the spatio-temporal simulation study, using a latent field configuration ($\sigma_1 = 1$, $\sigma_2 = 0.5$, $\sigma_3 = 0.3$) and a temporal coefficient $ a_2 = 0.5$.}
         \label{fig:st_latent_fields_y_2}
\end{figure}

\begin{figure}
         \centering
         \includegraphics[trim={0cm 4cm 0cm 1cm},clip,width=\textwidth]{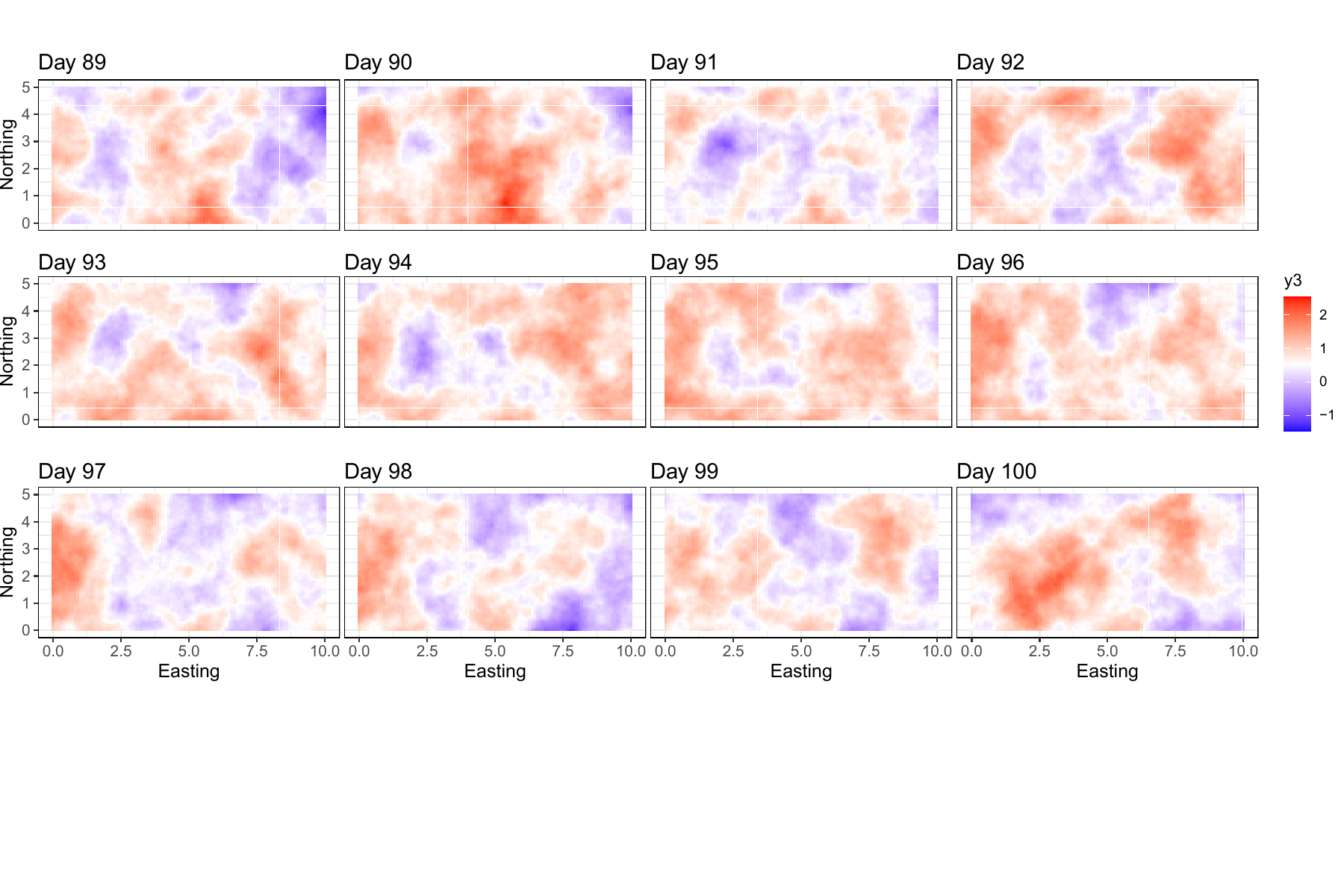}
         \caption{Final 12-day sequence of simulated latent fields for simulated soil moisture $y_3$ in the spatio-temporal simulation study, using a latent field configuration ($\sigma_1 = 1$, $\sigma_2 = 0.5$, $\sigma_3 = 0.3$) and a temporal coefficient $a_3 = 0.6$.}
         \label{fig:st_latent_fields_y_3}
\end{figure}

\begin{figure}
         \centering
         \includegraphics[trim={0cm 7cm 0cm 1cm},clip,width=\textwidth]{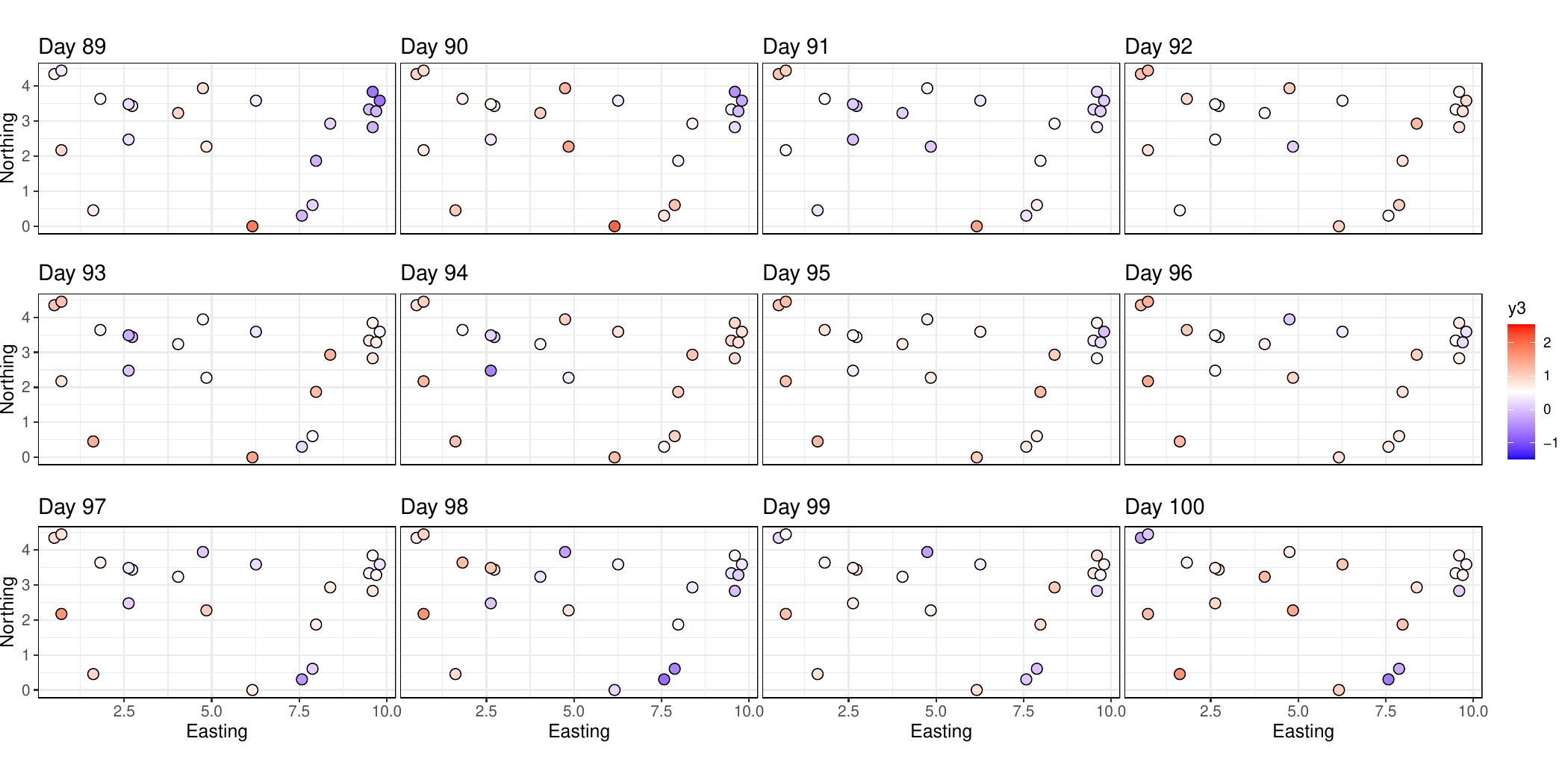}
         \caption{Final 12-day of simulated point soil moisture data for $y_3$ in the spatio-temporal simulation study. Twenty-two points are randomly selected from the realisation surface of simulated soil moisture $y_3$ on each day.}
         \label{fig: st_point}
\end{figure}

\begin{figure}
         \centering
         \includegraphics[trim={0cm 7cm 0cm 0cm},clip,width=\textwidth]{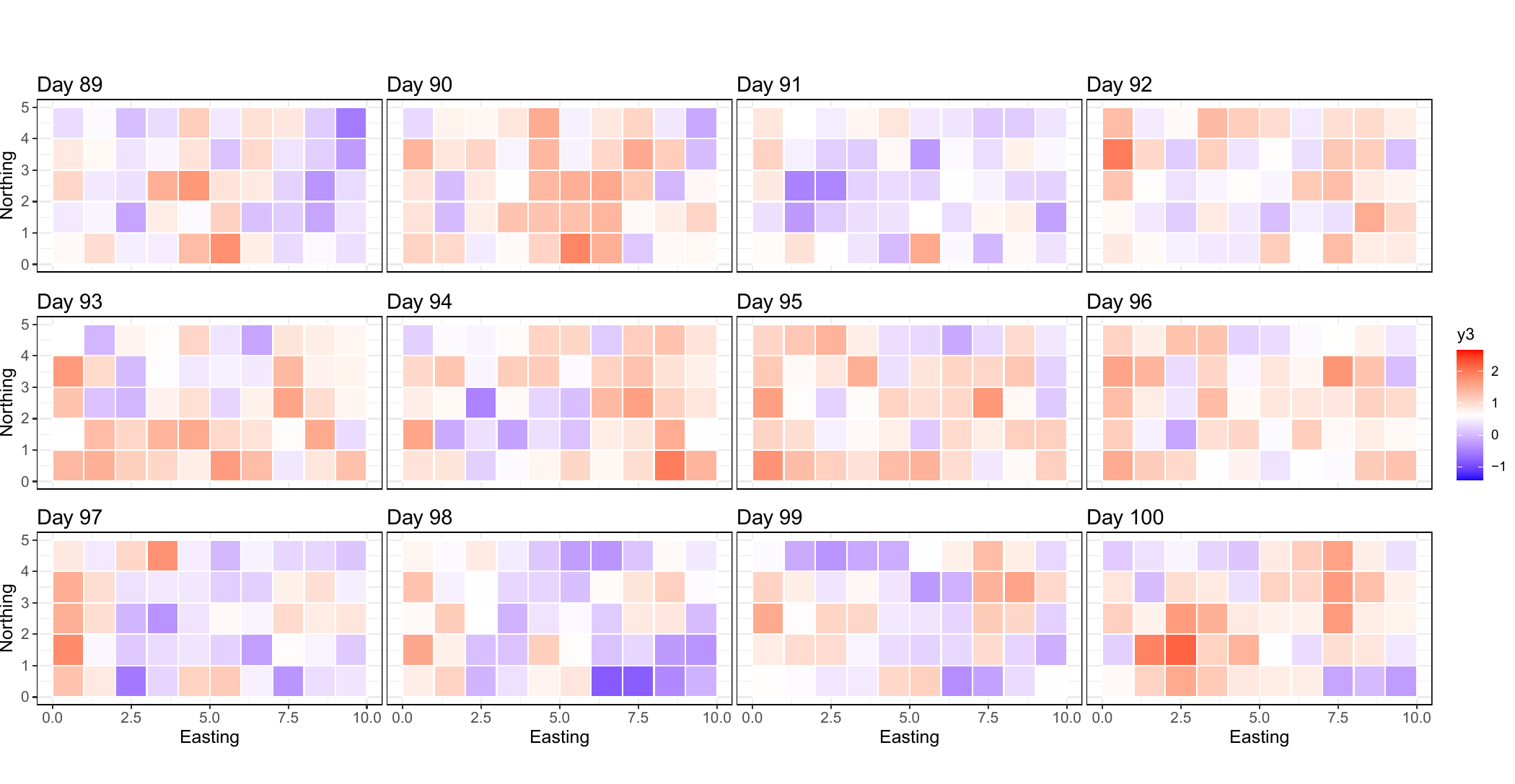}
         \caption{Final 12-day of simulated grid soil moisture data for $y_3$ in the spatio-temporal simulation study. It is averaged by 10,000 points generated from the realisation surface of simulated soil moisture $y_3$ on each day.}
         \label{fig: st_grid}
\end{figure}

The latent field $\eta_3$ simulates soil moisture dynamics using a spatially continuous and temporally autoregressive dependence structure. The spatial field is modelled as a Gaussian process with a Matérn covariance (smoothness $\rho_3=2$, variance $\sigma_3=0.1$), and the temporal dynamics as an AR(1) process with coefficients $a_1=0.4$, $a_2=0.5$, and $a_3=0.6$ corresponding to rainfall, soil moisture, and VWC, respectively. The autoregressive parameters are chosen based on the exploratory analysis of the real sensor data, which reflects the characteristics of the real data. The simulation fields span over a 5 $\times$ 10 spatial domain and evolve over 100 simulated time points, but only the final 12 are shown here.

Both the point data and grid data, along with the latent fields, are simulated by the models \eqref{eq:st_fusion_point_model} and \eqref{eq:st_fusion_grid_model} according to the workflow in Figure \ref{fig:st_flow_process}. The sensor network is simulated by sampling 22 locations across all time points. At each time point, we sample 10,000 values to evaluate the Matérn Gaussian random field, and average the values within each grid cell to approximate the spatial integral. After the independent daily data are generated, the AR(1) process is used to introduce the temporal dependence across time into the data.

The point data include a measurement error denoted as $\tau_{p_3}$. In contrast, the grid data includes a measurement error denoted as $\tau_{g_3}$, with $\tau_{g_3}$ being greater than $\tau_{p_3}$ because the sensor data is considered to be more accurate. This indicates that the grid data has greater measurement uncertainty compared to the point data, which aligns with the real-world data characteristic, where sensor data are generally more accurate than satellite image data.

Figure \ref{fig:st_latent_fields_y_1}, \ref{fig:st_latent_fields_y_2} and \ref{fig:st_latent_fields_y_3} demonstrate how the latent fields change across space and over time. Figure \ref{fig: st_point} shows the spatial distribution of sparse sensor locations, and Figure \ref{fig: st_grid} shows the coarse-resolution satellite data covering the whole study catchment. The autoregressive coefficients of each variable are $a_1=0.4$, $a_2=0.5$ and $a_3=0.6$ respectively, which is quite moderate temporal dependence, so the spatial pattern across different time points is not very obvious. Since the point data and grid data are a combination of all three latent fields ($\mu_1$, $\mu_2$, and $\mu_3$), it is even challenging to tell the temporal trends in the point data and grid data figures.

The prior distributions' parameters in the spatio-only and spatio-temporal model, such as intercepts, scaling parameters, and spatial parameters, are the same as those used in the spatio-only model, except for the temporal coefficients, whose priors are specified separately in Table \ref{tab: Priors specification for st_moint model parameters}. The penalised-complexity (PC) priors are used here for the temporal coefficients to guide the Bayesian inference process towards less complex solutions by penalising complexity and the distance from the base model by shrinking the range toward infinity and the marginal variance toward zero \citep{fuglstad2019constructing}.

\begin{table}
\centering
\caption{Prior specification for the temporal coefficient in the spatio-temporal data fusion model.}
\begin{tabular}{rrr}
  \hline
 Param & Informative prior  & Non-informative prior  \\ 
  \hline
    $\alpha_1$ &  & N(0,10)  \\ 
  $\alpha_2$ &  & N(0,10)  \\ 
  $\alpha_3$ &  & N(0,10)  \\ 
  $\beta_1$ &  & N(0,10)  \\ 
  $\beta_2$ &  & N(0,10)  \\ 
  $\theta_1$ &  & N(0,10)  \\ 
  $\rho_1$ & $\mathrm{PC}\left(\rho_0, \alpha\right)$ &   \\ 
  $\rho_2$          & $\mathrm{PC}\left(\rho_0, \alpha\right)$&   \\ 
  $\rho_3$          & $\mathrm{PC}\left(\rho_0, \alpha\right)$ &   \\ 
  $\sigma^2_1$      & $\mathrm{PC}\left(\sigma_0, \alpha\right)$&   \\ 
  $\sigma^2_2$      & $\mathrm{PC}\left(\sigma_0, \alpha\right)$ &   \\ 
  $\sigma^2_3$      & $\mathrm{PC}\left(\sigma_0, \alpha\right)$ &   \\ 
$\sigma^2_{e_1}$    &  & $\mathrm{PC}\left(\tau_{e_1}, \alpha\right)$   \\ 
  $\sigma^2_{e_2}$  &  & $\mathrm{PC}\left(\tau_{e_2}, \alpha\right)$   \\ 
  $\sigma^2_{e_3}$  &  & $\mathrm{PC}\left(\tau_{e_3}, \alpha\right)$ \\ 
  $\sigma^2_{e_1}$    &  & $\mathrm{PC}\left(\tau_{e_1}, \alpha\right)$   \\ 
  $\sigma^2_{e_2}$  &  & $\mathrm{PC}\left(\tau_{e_2}, \alpha\right)$   \\ 
  $\sigma^2_{e_3}$  &  & $\mathrm{PC}\left(\tau_{e_3}, \alpha\right)$ \\ 
  $a_1$    &  & $\mathrm{PC}\left(0.5, \alpha\right)$   \\ 
  $a_2$  &  & $\mathrm{PC}\left(0.5, \alpha\right)$   \\ 
  $a_3$  &  & $\mathrm{PC}\left(0.5, \alpha\right)$ \\ 
   \hline
\end{tabular}
\label{tab: Priors specification for st_moint model parameters}
\end{table}

\bibliography{bibliography}
\bibliographystyle{elsarticle-harv}

\end{document}